\documentclass[defended, 11pt]{new_cit_thesis}
\usepackage{graphicx}
\usepackage{amsmath}
\usepackage{amssymb}
\usepackage{lscape}

\newcommand{\PSbox}[3]{\mbox{\rule{0in}{#3}
\includegraphics{#1}\hspace{#2}}}
\newcommand{\nn}{\nonumber}
\def\xslash#1{{\rlap{$#1$}/}}
\providecommand{\openone}{\leavevmode\hbox{\small1\kern-3.8pt\normalsize1}}

\degreeaward{Doctor of Philosophy}
\author{Margaret E. Wessling}
\title{Heavy Pentaquarks in the Diquark Model and the Large $N_c$ Expansion}                 
\university{California Institute of Technology}    
\address{Pasadena, California}                     
\unilogo{cit_logo}                              
\copyyear{2005} 
\date{May 16, 2005}                           
\pubnum{68-2561}                                       
\begin{document}

\maketitle

\begin{frontmatter}
\makecopyright
\acknowledgements

I would like to thank my advisor, Mark Wise, for everything he has taught me over the past five years.  I would also like to thank Iain Stewart for a rewarding collaboration that produced the work presented here in Chapter 2, and Aneesh Manohar and Elizabeth Jenkins for useful discussions and advice on pentaquarks in large $N_c$.  

In addition, I would like to thank my family and friends for keeping me sane throughout the thesis-writing process.  I thank Matt Matuszewski for help with Mathematica and for all his encouragement during these past few weeks.  Thanks to Carlos Mochon for many useless but entertaining conversations on the fourth floor of Lauritsen.  Finally, thanks to my parents, John and Julie Wessling, and my brother Pete Wessling, for their love and support.

\begin{abstract}

Recent experimental evidence for the $\Theta^+(1540)$ has given rise to much theoretical interest in exotic baryons.  The $\Theta^+$ is a baryon that has strangeness $S=+1$, meaning that it contains an anti-strange quark.  Thus it cannot be constructed from three quarks, unlike all other known baryons; it needs at least an extra quark-antiquark pair. It is usually modeled as a pentaquark state in the ${\bf \overline{10}}$ representation of flavor $SU(3)$, with flavor content $\bar{s}uudd$. 
 
This thesis considers possible heavy pentaquarks, in which the antiquark is charmed or bottom rather than strange.   In the context of the diquark model of Jaffe and Wilczek, it is argued that negative-parity pentaquarks of this type may be lighter than their positive-parity counterparts, and hence are likely to be stable against strong decay.  Estimates are made for their masses, and their weak decays are discussed.  Isospin relations are found between the decay rates for different possible decay channels.

Negative-parity heavy pentaquarks are also considered in a less model-dependent way, in the context of a $1/N_c$ expansion, where $N_c$ is the number of colors. Heavy quark effective theory is also employed.  Mass relations are found between the mass splittings of heavy pentaquarks and those of nonexotic baryons, and $SU(3)$-breaking corrections to these relations are computed.   The results could be helpful in interpreting experimental data if heavy pentaquarks are observed.

\end{abstract}
\tableofcontents
\listoffigures
\listoftables
\end{frontmatter}

\chapter{Introduction}

The structure of hadrons is surprisingly poorly understood.  The Standard Model gives a detailed description of the couplings of quarks and gluons, and at very high energy scales, $>>\Lambda_{QCD}$, the strong coupling constant $\alpha_s$ is small enough that a perturbative expansion is useful.  At very low energies, one can apply chiral perturbation theory, essentially considering hadrons to be fundamental particles and ignoring their constituent quarks.  However, at intermediate energies around $\Lambda_{QCD}$--corresponding to a length scale around the size of a hadron--the strong coupling constant becomes large, and yet the quark structure is too important to be ignored.  A hadron is full of ``brown muck" consisting of gluons and quark-antiquark pairs popping in and out of existence; this muck tends to obscure our view of the hadron's interior.  Thus in order to understand what is going on inside a hadron, we are reduced to developing phenomenological quark models, to using abstract techniques like the large $N_c$ expansion, or to doing numerical calculations on the lattice.

In order to refine our understanding of hadrons, it is helpful to have as much empirical data as possible.  It is particularly useful to be able to observe the behavior of quarks in new and different environments, in bound states of unusual composition.  That is one reason why the prospect of exotic hadrons has been so intriguing.

\section{The $\Theta^+(1540)$}
\subsection{History}

Since the early days of the quark model, there has been speculation about exotic hadronic states.  A meson consists of a quark plus an antiquark, a baryon of three quarks; why not, say, a tetraquark with two quarks and two antiquarks, or a dibaryon with six quarks?  Of course, atomic nuclei can contain large numbers of quarks, but they behave, to a good approximation, as collections of three-quark nucleons, and are not considered to be ``exotic'' in this sense.

One possibility that received a fair amount of attention was the pentaquark, an exotic baryon containing four quarks and an antiquark.  (The name ``baryon'' is appropriate because such a state has baryon number one, although it is not made up of the usual three quarks.)  In the 1980's, experimental searches were made for pentaquarks, and tentative listings appeared in the Particle Data Book \cite{pdg82}, only to disappear shortly thereafter. (For a review of the history of exotic baryon searches, see \cite{maltman}.)  In 1988, the Particle Data Group announced, (\cite{pdg88}, quoted in \cite{jaffexotica}),

\bigskip

``The general prejudice against baryons not made of three quarks and the lack of any experimental activity in this area make it likely that it will be another 15 years before this issue is decided.''

\bigskip

\noindent This statement was to prove prophetic.
  
It was shown in the mid-'80's that the chiral soliton model for baryons \cite{chiralsoliton} gives rise to exotic $SU(3)$ flavor multiplets, such as the ${\bf \overline{10}}$ and the ${\bf 27}$, whose members have quantum numbers that cannot be produced by any combination of three quarks.  However, these were generally dismissed as unphysical until 1997, when Diakonov, Petrov, and Polyakov used the model to predict a particle they named the $Z^+$ \cite{diakonov}.  The $Z^+$ was a baryon in the ${\bf \overline{10}}$, with minimal quark content $\bar{s}udud$.  Diakonov, {\it et al.} predicted a mass of 1530 MeV and a strikingly narrow decay width, less than 15 MeV, for the $Z^+$.   

In 2003, exactly 15 years after the Particle Data Group statement above, the LEPS collaboration \cite{nakano} experimentally detected a baryon that appeared to be nothing other than the $Z^+$.  It decayed to $NK$, meaning that it had positive strangeness ({\it i.e.}, contained an anti-strange quark), and thus could not be constructed from just three quarks.  It had a mass of 1540 MeV and a width below the resolution of the detector, less than 25 MeV.   Many other experiments \cite{theta} soon went on to observe this exotic baryon, now renamed the $\Theta^+$ and thought by many to be a pentaquark.\footnote{Some authors \cite{bicudo} have suggested that the $\Theta^+$ may in fact be a {\it hepta}quark, a bound state of $K$, $N$, and $\pi$.  This view will not be discussed further here.}

\subsection{The ${\bf \overline{10}}$ Multiplet}

%%%%%%%%%%%%
\begin{figure}
\PSbox{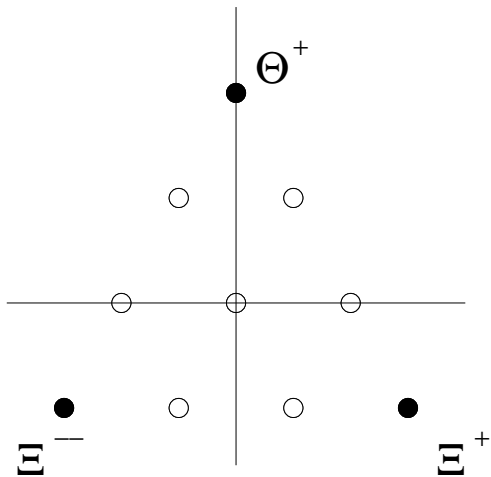 hoffset=40 voffset=20 hscale=80
vscale=80}{3.0in}{3.0in}
\caption[The ${\bf \overline{10}}$ multiplet]{The ${\bf \overline{10}}$ multiplet of $SU(3)_f$.  The vertical axis represents hypercharge, $Y$; the horizontal axis represents the third component of isospin, $I_3$.  The three states at the vertices of the triangle have exotic quantum numbers; the white circles represent cryptoexotic states.  
}
\label{fig:tenbar}
\end{figure}
%%%%%%%%%%%%%%
Given its flavor content, the $\Theta^+$ could have isospin 0, 1, or 2.  However, no isospin partners have been detected, supporting the interpretation that it has $I=0$ and is at the tip of the ${\bf \overline{10}}$ $SU(3)_f$ multiplet (Figure \ref{fig:tenbar}), as predicted by Diakonov, {\it et al.}.  The three states at the vertices of the triangle, shown in black in the figure, have quantum numbers that are truly exotic: they could not arise from a combination of just three quarks.  In addition to the $\Theta^+$ at the top of the diagram, there has been some experimental evidence for the state in the bottom left corner, with valence quark content $\bar{u}dsds$. The NA49 Collaboration \cite{alt} observed a mass of 1862 MeV for this state, which it called $\Xi_{3/2}^{--}$; it is listed in the 2004 Particle Data Book \cite{pdg} as the $\Phi(1860)$, at the one-star confidence level.  Other experiments \cite{noxi} have reported null results in searches for the $\Xi/\Phi$ pentaquark.   

The states shown in white in Figure \ref{fig:tenbar} are known as ``cryptoexotic'' states.  Each contains a quark and an antiquark of the same flavor--for example, $\bar{u}usud$ or $\bar{d}dsus$.  This results in quantum numbers that are indistinguishable from those of an ordinary three-quark baryon.  Thus, experimenters may already have observed some of these states without realizing that they were part of a pentaquark multiplet.  The designers of quark models explaining the $\Theta^+$ have explored the possibility that a number of known baryons may be members of the ${\bf \overline{10}}$ \cite{maltman, JaffeWilczek}.

\section{Experimental Questions}

\subsection{Is it real?}

The $\Theta^+$ is listed in the 2004 Particle Data Book \cite{pdg} with three stars. It should be noted, however, that a significant number of experiments have searched for the $\Theta^+$ without success \cite{notheta}, casting doubt on its existence.  Much speculation has occurred as to how the $\Theta^+$ might avoid detection in some experiments while being seen clearly in others.  There are some suggestive differences between the experiments that see the $\Theta^+$ and those that do not. Many of the experiments with positive results involve photoproduction, while those with negative results use either $e^+e^-$ production or high-energy proton collisions \cite{close, hicks}.  Perhaps the production of the $\Theta^+$ is suppressed in some mechanisms and enhanced in others.  Hicks \cite{hicks} points out that, in the case of  $e^+e^-$ annihilation, the production of (three-quark) baryons occurs at a lower rate than that of mesons, and suggests that the rate may be still lower for pentaquarks.  Karliner and Lipkin \cite{kl} suggest that the $\Theta^+$ may arise only via a particular production mechanism, such as the decay of a cryptoexotic $N^*$ resonance.  It is, of course, possible that the $\Theta^+$ may turn out not to exist, though it appears somewhat unlikely that so many positive results, from different experiments looking at different types of reactions, could all be explained away.  (For a table showing all the experiments and the reactions they observed, see \cite{hicks}.)  On the other hand, it should be noted that most of the experiments reporting positive results have low statistics, while those with negative results have higher statistics.  Clearly, more experimental evidence is needed; new high-statistics experiments that are planned or already under way at CLAS and COSY-TOF will likely settle the question within the next year or so \cite{hicks2}.  

Quite recently, Jefferson Lab issued a press release announcing the results of the latest experiment at CLAS \cite{jlab}.  Observing the same reaction in which the $\Theta^+$ was previously detected at SAPHIR ($\gamma p\to K^+\bar{K}^0n$), with much higher statistics, CLAS found no signal.  This result casts a great deal of doubt on the existence of pentaquarks, although it does not represent definitive proof that the $\Theta$ is not real.  As of this writing, the results are not yet in from JLab's g10 experiment, which uses a deuterium target to reproduce the original CLAS experiment that reported positive results, with higher statistics.  A recent response to JLab's announcement \cite{Nam} suggests that the $\gamma p$ reaction at CLAS may have produced a null result because of an asymmetry in the interactions of the proton and the neutron; however, this suppression mechanism works much better if the $\Theta^+$ has spin $3/2$, which appears unlikely.

In this thesis, we will take an agnostic position as to the reality of the $\Theta^+$.  None of the models we investigate can make a definitive statement about whether or not pentaquarks exist; however, given the assumption that the measurements of the $\Theta^+$ are real, these models allow us to extrapolate the properties of other types of pentaquarks.  If the $\Theta^+$ does turn out to exist, theorists will face the problem of explaining why it fails to appear in certain experiments.  On the other hand, if the consensus turns out to be that the $\Theta^+$ measurements were false, we will face a different puzzle: why {\it don't} pentaquarks exist?  QCD as we currently understand it certainly allows for them; could there be some hidden aspect of hadron dynamics that rules them out?   Still another possibility is that there is no exotic baryon with the mass and width of the $\Theta^+$, but that real pentaquarks are lurking somewhere else in the hadron spectrum and have yet to be discovered. 

\subsection{Width and Parity}

Nearly all the experiments that have found evidence for the $\Theta^+$ have reported that its width is below the resolution of their detectors, {\it i.e.}, $<10-20$ MeV, although ZEUS and HERMES give the width as $8 \pm 4$ and $13 \pm 9$ MeV respectively \cite{hicks}.  Indirect upper limits based on elestic scattering put the width even lower, at a few MeV or even $<1$ MeV \cite{narrow}.

This narrow width is unexpected and intriguing.  In an uncorrelated quark model, the width would be several hundred MeV; it should be easy for a pentaquark to fall apart into a baryon and a meson.  Follow-up work in the chiral soliton model after the original paper by Diakonov, {\it et al.} predicted a width of $\sim100$ MeV \cite{weigel}.  

There have been several correlated quark model proposals attempting to account for the narrowness of the $\Theta^+$, including the diquark model \cite{JaffeWilczek} and the triquark model \cite{karliner}, which will be discussed further in the next chapter.  The basic idea is that it would be relatively difficult to rearrange specially correlated groups of quarks into the final state $KN$, and hence the decay would proceed slowly.  It has also been argued that a narrow decay width arises naturally in a QCD sum rule approach \cite{wang}.

 The parity and spin of the $\Theta^+$ have not been determined experimentally.  Most researchers assume that the spin is $\frac{1}{2}$; it is unlikely that the ground state could have a higher spin.  The parity, however, is somewhat controversial.  The diquark and triquark pictures require positive parity, as does the chiral soliton model.  It has been argued \cite{carlson} that a positive-parity pentaquark should always have a very narrow decay width, regardless of specific dynamical assumptions, because a positive-parity wavefunction would have quite a small overlap with the wavefunction of the final state $K+N$.  Positive parity thus looks attractive because it would help explain the oddly narrow width of the $\Theta^+$.  

On the other hand, studies using QCD sum rules \cite{sumrules} have indicated negative parity.  So have some lattice QCD studies \cite{lattice}, including a lattice calculation with operators based on the diquark model \cite{sasaki}.  (Other lattice searches \cite{nolattice} have found no bound pentaquark state at all.) There are a number of technical issues that call into question the reliability of studies of these two types, as discussed in \cite{maltman}.  In the case of the QCD sum rules, some important OPE contributions may have been neglected.  In the lattice studies, the quark masses used are much heavier than the physical values, and additional work at lighter quark masses might change the results; there is also a question of whether the appropriate operator has been chosen to represent the pentaquark.  In any case, the parity of the $\Theta^+$ remains an important experimental question.  There have been several suggestions of ways to measure the parity, by looking at the patterns of pentaquarks' strong decays and at their production mechanisms \cite{parity}.  

\section{Heavy Pentaquarks}

Now that there is evidence for the $\Theta^+$, it is natural to consider the possibility of other exotic baryons as well.  The various models that have been used to explain the properties of the $\Theta^+$ provide a framework in which to analyze different types of pentaquarks, including those beyond the ${\bf \overline{10}}$ multiplet.  In this thesis, we will explore the possibility of heavy pentaquarks, {\it i.e.}, those containing charm or bottom quarks.\footnote{One might imagine that the methods used here could easily be extended to pentaquarks containing top quarks as well.  However, the top quark decays much more quickly than the time scale for hadron formation.  Its decay width may be estimated by $\frac{m_{top}g_2^2}{8\pi} \sim$ 1 GeV.}  There is some (highly controversial) experimental evidence for a charmed pentaquark \cite{charm}, which will be discussed further in the next chapter.

Chapter 2 considers the implications of Jaffe and Wilczek's diquark model for heavy pentaquarks.  We argue that, due to Fermi and Bose statistics, negative-parity heavy pentaquarks should be lighter than their positive-parity counterparts in the diquark picture.  We predict a triplet of such states, estimate their masses, and show that they are likely to be stable against strong decay, making them potentially accessible to experimental searches.  Their possible weak decays are analyzed, and isospin and $SU(3)$ relations are found for many of the decay rates.

Chapter 3 considers negative-parity heavy pentaquarks in a broader context, using the large $N_c$ expansion.  This allows some properties of the pentaquark states to be derived without the need for any particular dynamical assumptions.  Heavy quark symmetry also comes into play, allowing additional relations to be predicted.  We relate the masses of various heavy pentaquark multiplets to one another, and also to nonexotic heavy and light baryons, and we consider $SU(3)$ breaking effects.

Appendices A and B contain figures and tables related to the material in Chapter 3.  Appendix C turns away from the subject of pentaquarks; it contains earlier research on gravity in extra dimensions \cite{strings}.  We calculate the gravitational potential energy between infinitely long parallel strings with tensions $\tau_1$ and $\tau_2$. Classically, it vanishes, but at one loop, we find that the long range gravitational potential
energy per unit length is $U/L=  24G_N^2\tau_1\tau_2/(5 \pi a^2)$ + ..., where $a$ is the separation between the strings, $G_N$ is Newton's constant, and we set $\hbar =c=1$. The ellipses represent terms suppressed by more powers of $G_N \tau_i$. Typically, massless bulk fields give rise at one loop to a long range potential between $p$-branes in space-times of dimension $p+2+1$. The contribution to this potential from bulk scalars is computed for arbitrary $p$ (strings correspond to $p=1$) and in the case of three-branes its possible relevance for cosmological quintessence is commented on.

\chapter{Stable Heavy Pentaquarks in the Diquark Model}

\section{Correlated Quark Models and Positive-Parity Pentaquarks}

As mentioned above, in order to help explain the narrow width of the $\Theta^+$, a number of correlated quark models have been been proposed.  These include the diquark model of Jaffe and Wilczek \cite{JaffeWilczek} and the triquark model of Karliner and Lipkin \cite{karliner}.  Each of these posits that the $\Theta$ pentaquark has orbital angular momentum $\ell=1$, and thus positive parity.  Such a configuration should lead to a narrow width because of its small overlap with the conventional $KN$ state into which the $\Theta$ decays, as argued in \cite{carlson}.

In the triquark model, the $\Theta^+$ is pictured as a $ud$ diquark plus a $ud\bar{s}$ triquark.  The two quarks in the diquark are combined antisymmetrically; it has spin 0 and is in a ${\bf \bar{3}}$ of color and a ${\bf \bar{3}}$ of flavor.  The $u$ and $d$ within the triquark are combined symmetrically, into a state of spin 1, color ${\bf 6}$, and flavor ${\bf \bar{3}}$.  They combine with the $\bar{s}$ to form a color triplet triquark with spin 1/2 and flavor ${\bf \bar{6}}$.  Karliner and Lipkin argue that there must be a relative $P$-wave between the diquark and triquark, because otherwise repulsive hyperfine interactions between the two clusters would break up the pentaquark into $K+N$.  With the $P$-wave, the angular momentum barrier separates the triquark and diquark and prevents such color-magnetic interactions between them. 

\begin{figure}
\PSbox{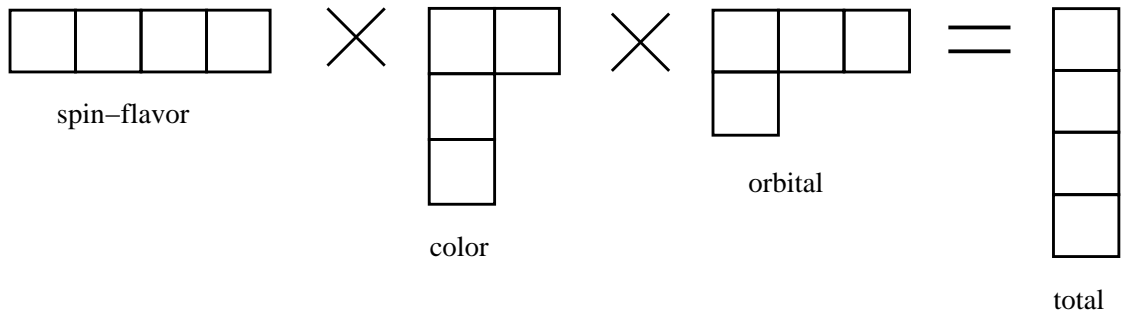 hoffset=20 voffset=20 hscale=40
vscale=40}{6.8in}{2.2in}
\caption[Young tableaux for the four quarks in the $\Theta^+$]{Young tableaux for the four quarks in the $\Theta^+$.  The spin-flavor part of the wavefunction is completely symmetric, while the color part has mixed symmetry.  In order to get a completely antisymmetric total wavefunction, there must be some orbital antisymmetry, provided by a $P$-wave.}
\label{fig:posparity}
\end{figure}

In the diquark model, the $\Theta^+$ is divided into two $ud$ diquarks, each in a state of spin 0, ${\bf \bar{3}}$ color, and ${\bf \bar{3}}$ of flavor, plus a single $\bar{s}$ antiquark.  If we ignore color and approximate the diquarks as point particles, the two are identical bosons; thus their wavefunction must be symmetric in spin and flavor in order to obey Bose statistics.  However, the overall wavefunction of the $\Theta$ must be completely antisymmetric according to Fermi statistics.  We must insert a unit of orbital angular momentum between the two diquarks in order to provide the necessary antisymmetry, as illustrated in Figure \ref{fig:posparity}.
%%%%%%%%%%%%
%%%%%%%%%%%%%%
 In this way, the P-wave structure arises naturally within the diquark model.

Both groups have considered possible heavy analogues of the $\Theta$: the $\Theta_c$ (with the $\bar{s}$ replaced by a $\bar{c}$), and the $\Theta_b$ (with a $\bar{b}$).  Jaffe and Wilczek made a simple mass estimate suggesting that these states might be stable against strong decay \cite{JaffeWilczek}.  Their results, 
%%%%%%%%%%%%%%%
\begin{align}
m_{\Theta_c} & \simeq 2710\,{\rm MeV} \nn \\   
m_{\Theta_b} & \simeq 6050\,{\rm MeV},
\end{align}
%%%%%%%%%%%%%%%%
are respectively $100\,{\rm MeV}$ and $165\,{\rm MeV}$ below the strong decay thresholds to $pD^-$ and $n B^+$.  This would be quite interesting if true: states that are stable against strong decay are likely to be long-lived and hence relatively easy to find experimentally.  However, given the uncertainties involved, and the fact that these masses are rather close to threshold, such a conclusion is questionable.  

Other models give other results: using their triquark model, Karliner and Lipkin estimated \cite{karlinercharm}
%%%%%%%%%%%%%%%
\begin{align}
m_{\Theta_c} & \simeq 2985\,{\rm MeV} \nn \\   
m_{\Theta_b} & \simeq 6398\,{\rm MeV},
\end{align}
%%%%%%%%%%%%%%%%
putting both states above the strong decay threshold.  Estimates based on a constituent quark model \cite{cheung} also fail to support stability against strong decay, and so does a lattice QCD calculation with operators inspired by the diquark model \cite{sasaki}.  On the other hand, a calculation involving so-called tensor diquarks (spin 1, flavor {\bf 6}) \cite{shuryak} gives masses close to those found by Jaffe and Wilczek \cite{huang}. 

The H1 Collaboration has published a claim for the detection of a charmed pentaquark \cite{charm}.  The mass they observe, 3099 MeV, is higher than any of the estimates above.  Some authors \cite{nowak} have suggested that this state, if indeed it exists, may be an excited state of the $\Theta_c$.  Other experiments have searched for the same particle and found nothing \cite{nocharm}.  

\section{Negative-Parity Heavy Pentaquarks}

As discussed in a recent work by the author and collaborators \cite{stewart}, the diquark picture suggests a way of constructing heavy pentaquarks that are more likely to be stable against strong decay.  The idea is straightforward: with a heavy antiquark, it is possible to make an exotic state in which the two diquarks have different flavor content.  Diquark pairs of $u$, $d$, and $s$ quarks come in three types: $(ud)$, $(ds)$, and $(su)$; as noted above, these form antitriplets with respect to $SU(3)$ flavor and $SU(3)$ color.  We will denote the interpolating fields for these diquarks by
%%%%%%%%%%%%%%
\begin{equation} \label{notation}
  \phi^{a \alpha} = \epsilon^{abc}\epsilon^{\alpha \beta \gamma}
    q_{b\beta}q_{c\gamma} \,,
\end{equation}
%%%%%%%%%%%%%%%%
where $\alpha, \beta$, and $\gamma$ are $SU(3)$ color triplet indices and $a,b$,
and $c$ $SU(3)$ flavor triplet indices ({\it i.e.}, for $q_a$, $a=1$ corresponds to an up quark, $a=2$ to a down quark, and $a=3$ to a strange quark). 

\begin{figure}
\PSbox{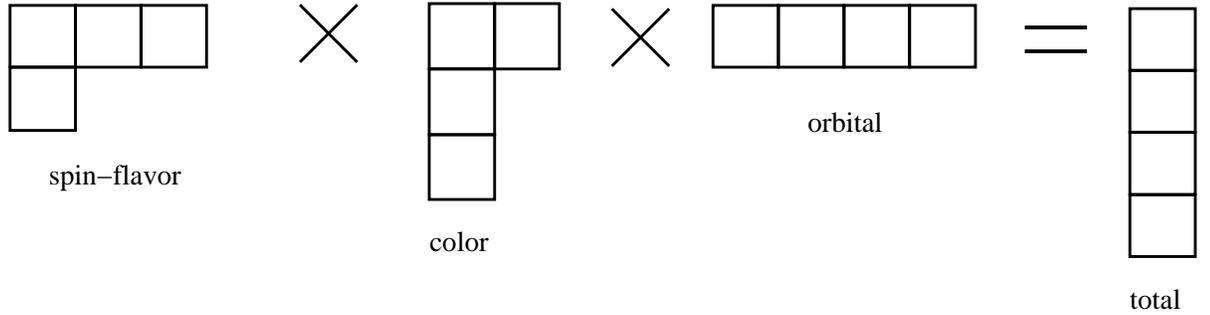 hoffset=20 voffset=20 hscale=40
vscale=40}{6.8in}{2.2in}
\caption[Young tableaux for the four light quarks in a negative-parity heavy pentaquark]{Young tableaux for the four light quarks in a negative-parity heavy pentaquark.  The spin-flavor part of the wavefunction is no longer completely symmetric, so the two diquarks may be in an $S$-wave.} 
\label{fig:negparity}
\end{figure}

In the $\Theta^+(1540)$, the two diquarks are in the ${\bf \overline{6}}$ of $SU(3)$ flavor, and the pentaquark itself is in the ${\bf \overline{10}}$, as mentioned above.  In this notation, it looks like
%%%%%%%%%%%%%%
\begin{equation}
\bar s^\alpha \epsilon_{\alpha\beta \gamma}\phi^{3\beta}\phi^{3\gamma}, 
\end{equation}
%%%%%%%%%%%%%%%%
and the $\Theta_{b,c}$, in the ${\bf \overline{6}}$ of $SU(3)_f$, similarly become
%%%%%%%%%%%%%%
\begin{equation}
\bar Q^\alpha \epsilon_{\alpha\beta \gamma}\phi^{3\beta}\phi^{3\gamma}.
\end{equation}
%%%%%%%%%%%%%%%%
Here we consider heavy pentaquarks of another type:
%%%%%%%%%%%
\begin{align}
T_a  & =\epsilon_{abc}\,\epsilon_{\alpha \beta \gamma}\, \bar c^{\alpha}\, 
    \phi^{b \beta}\phi^{c \gamma}\,, \nn \\
R_a  & =\epsilon_{abc}\,\epsilon_{\alpha \beta \gamma}\, \bar b^{\alpha}\, 
    \phi^{b \beta}\phi^{c \gamma}\,.
\end{align}
%%%%%%%%%%%
In these states, the two diquarks are no longer identical bosons; the spin-flavor part of their wavefunction no longer needs to be completely symmetric.  This eliminates the need to insert extra antisymmetry via orbital angular momentum (Figure \ref{fig:negparity}).  Such a pentaquark can be in an $\ell=0$ state, giving it negative overall parity.   The lack of $P$-wave excitation energy in these states suggests that they may be lighter than the $\Theta_c$ and $\Theta_b$. 

Note that the states we've defined here are indeed truly exotic, being baryons with charm $=-1$
(or beauty $=1$) and strangeness $=-1,-2$.\footnote{Strangeness and beauty, or bottomness,  are defined to be negative for an $s$ or $b$ quark, and positive for an $\bar{s}$ or $\bar{b}$ antiquark, while charm has the opposite sign convention.  This definition arose from a historical accident in the case of strangeness, but has been extended to other quarks based on their charge: those with positive charge follow the sign convention of charm, while those with negative charge follow the convention of strangeness.}   Consider, for example, the case in which the heavy antiquark is a $\bar{c}$: as far as flavor quantum numbers are concerned, we have $T_1=\bar c (ud)(su)$, $T_2=\bar c (ud)(sd)$ and $T_3=\bar c (su)(sd)$.    In contrast, the lighter $S$-wave analog of $T_a$ with $\bar c\to \bar s$ is cryptoexotic; it mixes
with excited nucleon states via annihilation and is therefore hard to detect.

To emphasize the strangeness and charge of the
states in this multiplet, we will often use the notation $T_1=T_s^0$,
$T_2=T_s^-$, $T_3=T_{ss}^-$ for charm and $R_1=R_s^+$, $R_1=R_s^0$,
$R_3=R_{ss}^0$ for bottom. Here $\{T_s^0,T_s^-\}$ and $\{R_s^+,R_s^0\}$ form
isospin doublets, while $T_{ss}^-$ and $R_{ss}^0$ are isospin singlets. 

%%%%%%%%%%%%
%%%%%%%%%%%%%

The possibility of exotic pentaquarks of the form $T_{1,2}$ was noted in 1987 \cite{Gignoux,Lipkin87}, and the E791 collaboration later performed an
experimental search \cite{Aitala:1997ja} in the mass range $2.75-2.91\,{\rm
  GeV}$, with a null result. (In Ref.~\cite{Lipkin87} the states $T_{1,2}$ were called $P_{\bar c s}$.)
In the context of the diquark model, the exotic ${\bf 3}$ multiplet $T_a$ with
the diquarks in a relative $S$-wave was first discussed by Cheung \cite{cheung},
but no mass estimate was given. Here we make mass estimates for $T_a$ and $R_a$
in the diquark picture. We point out that the $R_a$ and $T_a$ may be well below the strong
threshold, and that the $T_a$ could be $\sim 200\,{\rm MeV}$ lighter than the E791
search window.  The exotic nature of these states can be determined through
measurement of weak decays, and we devise isospin and $SU(3)$ relations which
could be used to  test their flavor quantum numbers further if they are observed.

\subsection{Mass Estimates}

For our purposes, the most important feature of the $T_a$ and $R_a$ states is
that, within the diquark picture, they are more likely to be stable against
strong decays than $\Theta_{c,b}$, since they do not require the excitation
energy associated with a $P$-wave, ${\cal U}_{P\!-\!wave}$. To make a rough
estimate of their masses, we write
%%%%%%%%%%%%%%%%%%%%
\begin{equation} \label{m1}
  m_{T_s}-m_{\Theta_c}= m_{R_s}-m_{\Theta_b}
  = \Delta_s -{\cal U}_{P\!-\!{\rm wave}},
\end{equation}
%%%%%%%%%%%%%%%%%%%%%%%%%%%%%
where $T_s$ and $R_s$ denote the isodoublet states. Here $\Delta_s$ is the
change in mass resulting from removing a $(ud)$ diquark and adding a $(us)$ or $(ds)$, while ${\cal U}_{P-{\rm wave}}$ is the energy associated with the $P$-wave.
$\Delta_s$ takes into account that the $T_s$ ($R_s$) contains a strange quark, unlike
the $\Theta_c$ ($\Theta_b$).  To estimate the extra mass contributed by the strange quark, we use
%%%%%%%%%%%%
\begin{align} \label{deltas}
\Delta_s \simeq m_{\Xi_c} - m_{\Lambda_c} = 184\,{\rm MeV} \,.
\end{align}
%%%%%%%%%%%%
Why is this a reasonable estimate of the mass difference?  There is evidence \cite{jaffe2} suggesting that many ordinary three-quark baryons, including nucleons and charmed baryons, are well-modeled as a diquark plus one extra quark.  Some of this evidence arises from the spectra of $uds$ and $udc$ baryons, which reflect the different binding energies of the antisymmetric (spin 0) and symmetric (spin 1) $ud$ diquarks.  More evidence comes from jet events producing $\Lambda$ and $\Sigma$ baryons.  The  $\Lambda$ and $\Sigma$ have the same flavor content, but the $\Lambda$ contains an antisymmetric $ud$ diquark, making it easier to assemble and hence more common than the $\Sigma$.

Still other indications came from deep inelastic scattering experiments with nucleons.  When the nucleons were bombarded with high-energy electrons, they sometimes behaved as though all their energy were concentrated in a single quark.  This happened four times as often with protons as with neutrons; the factor of four was conjectured to arise from the square of the quark's charge ($\left(\frac{2}{3}\right)^2$ for the up quark, $\left(-\frac{1}{3}\right)^2$ for the down quark).  But why did this never happen with the down quark in the proton or the up quark in the neutron?  The answer was conjectured to be that inside each nucleon, two quarks, an up and a down, are tied together in a tightly bound structure: a diquark.  Thus the diquark idea has a long history, and much supporting evidence, completely independent of pentaquarks.

In Eq. (\ref{deltas}), we look at the mass difference between $(us)c$ and $(ud)c$, where $(us)$, etc., represent spin 0 diquarks.  That is, in an environment containing a charm quark, we replace a strange diquark with a non-strange one.  This is just like going from $T_s$ to $\Theta_c$, aside from the angular momentum (which will be accounted for later) and the fact that another light diquark goes along for the ride.  Of course, there is a fair amount of uncertainty associated with the latter point; a five-quark environment may well produce a different mass than a three-quark one, and we stress that this is a rough estimate.  Nevertheless, it is a sensible approximation in light of our present knowledge about baryons.  (Incidentally, taking the difference between the masses of the $\Lambda$ and the proton, $(ud)s$ and $(ud)u$, does not change the estimate much: $\Delta_s\simeq m_{\Lambda}-m_p =177\,{\rm MeV}$.)

In a similar way, we estimate the $P$-wave energy by
%%%%%%%%%%%%%%
\begin{equation}
\label{pwave}
  {\cal U}_{P-\!{\rm wave}} \simeq m_{\Lambda'_{c}}-m_{\Lambda_c} 
    = 310\,{\rm MeV},
\end{equation}
%%%%%%%%%%%%%%%%%
where $\Lambda'_{c}$ denotes the excitation of the $\Lambda_c$ with $(ud)$ in a
$P$-wave relative to $c$; $m_{\Lambda'_{c}}=2594~{\rm MeV}$. This estimate is
supported by $P$-wave excitation energies for baryons built of light $u,d,s$
quarks; for example, $m_{\Lambda(1405)}-m_\Lambda = 291\,{\rm MeV}$.  Again, we assign a sizable uncertainty to the estimate in Eq. \eqref{pwave}; it is easily possible that it overestimates (or underestimates) the $P$-wave excitation energy for $\ell=1$ between two diquarks.  It also neglects possible Pauli-blocking effects between identical quarks in different diquarks.

Using Eqs. (\ref{m1}) and (\ref{pwave}) along with the Jaffe-Wilczek estimate,
$  m_{\Theta_c}=m_{\Theta}+m_{\Lambda_c}-m_{\Lambda} \simeq 2709~{\rm MeV}$,
gives the $T_s$ mass estimate 
%%%%%%%%%%%%%%%%%%%%%
\begin{equation} \label{masc} 
   m_{T_s} 
     \simeq  m_{\Theta}+m_{\Lambda_c}-m_{\Lambda} +m_{\Xi_c}-m_{\Lambda_c'}
    =  2580~{\rm MeV} \,.
\end{equation}
%%%%%%%%%%%%%%%%%%
For the strong decay $T_s\to D_s\, p$, the sum of the $D_s$ and proton masses is $2910~{\rm
  MeV}$; {\it i.e.}, eq. (\ref{masc}) puts the state $330\,{\rm MeV}$ below threshold
and $170\,{\rm MeV}$ below the E791~\cite{Aitala:1997ja} search region. 

A similar analysis in the case for which the heavy quark is a bottom quark gives the
mass formula
%%%%%%%%%%%%%%%
\begin{equation} \label{masb} 
  m_{R_s} 
  \simeq  m_{\Theta}+m_{\Lambda_b}-m_{\Lambda} +m_{\Xi_c}-m_{\Lambda_c'}
   =5920~{\rm MeV},
\end{equation}
%%%%%%%%%%%%%%%%%%%
which is $390\,{\rm MeV}$ less than the sum of the $B_s$ and proton masses.
Thus, even with the large uncertainties in the mass estimates in
Eqs. (\ref{masc}, \ref{masb}), it appears quite likely that these states are stable against strong decays.

The extra strange quark in $T_{ss}$ and $R_{ss}$ will
increase the mass of each by about $\Delta_s$ relative to $T_s$ and $R_s$
respectively, so that
%%%%%%%%%%%%% 
\begin{align} \label{masbcss}
 m_{T_{ss}} \simeq m_{T_s}+\Delta_s & = 2770\,{\rm MeV} \,, \\
 m_{R_{ss}} \simeq m_{R_s}+\Delta_s & = 6100\,{\rm MeV}\,.
\end{align}
%%%%%%%%%%%%%%%%
However, the presence of the extra strange quark does not make $T_{ss}$ and $R_{ss}$
closer to threshold: their strong decays must involve two $s$ quarks in the
final state, e.g.,  $R_{ss}\to B_s\,\Lambda$.

\subsection{Decays}

\begin{figure}
\PSbox{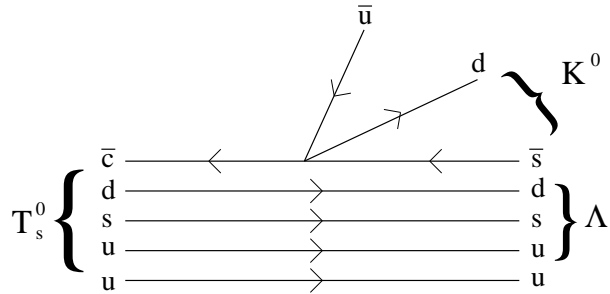 hoffset=50 voffset=20 hscale=50
vscale=50}{6.0in}{2.5in}
\caption[A possible nonleptonic decay]{A possible nonleptonic decay via the weak process $\bar c\to \bar s d\bar u$: $T_s^0\to \Lambda K^0$.  The $\bar{u}$ and the lower $u$ on the right hand side annihilate, leaving the $\Lambda$, $uds$, and the $K^0$, $\bar{s}d$.}
\label{fig:nonleptonic3}
\end{figure}

For the charmed and bottomed exotics, promising weak processes for detection are
$\bar c\to \bar s d\bar u$, $\bar b \to \bar c c \bar s$, and $\bar b\to \bar c
u\bar d$.  (See Figure \ref{fig:nonleptonic3} for an example.)  Typical nonleptonic decay channels such as $T_s^0\to p \phi \pi^-$
and $T_s^-\to \Lambda K^+\pi^-\pi^-$~\cite{Lipkin87} always break up the diquarks.
This may substantially decrease the corresponding partial widths, particularly
if the narrowness of the $\Theta^+(1540)$ is partly due to such an effect. For
bottom, this penalty can be postponed by decays to charmed exotics that preserve
the diquark correlation, for example, $\Theta_b\to\Theta_c\pi$~\cite{llsw3}. In
our case, the analogues are $R_a\to T_a \pi^+$ and $R_a\to T_a D^+$. Among the
two-body $\bar b\to \bar c u\bar d$ decays, these exotic-to-exotic channels are
also dynamically favored by factorization \cite{factor}, which favors producing
an energetic $u\bar d$ meson, and suppresses decays to energetic mesons built out
of other flavor combinations that appear in the decay.

%%%%%%%%%%%%
%%%%%%%%%%%%%

In order to search experimentally for these pentaquarks, it is important to consider decays in which the products can be easily detected.  In general, charged particles can be detected at, e.g., the Tevatron \cite{tevatron}, while neutral ones cannot; however, it is possible to infer the presence of a neutral particle that decays in a distinctive way to several charged particles.  For example, the $\rho^0$ can be reconstructed because it decays to two charged pions.

Assuming the $T_a$ and $R_a$ states decay weakly, there are several
promising discovery channels. For charm, the nonleptonic decays include
\begin{align} \label{Tlist}
  & T_s^0 \to \Lambda K^0,\, p\pi^-,\, p\phi\pi^-,\, \Lambda K^+\pi^-,\, 
   K^0 K^- p ,\, \phi K^0\Lambda ,\, K^0 K^+ \Xi^- \,,
   \\
  & T_s^- \to K^0\pi^- \Lambda,\, p\pi^-\pi^-,\, p \phi \pi^-\pi^- \,, \Lambda K^+\pi^-\pi^-,
   %K^0 K^0 \Xi^- 
   \nn \\
  & T_{ss}^- \to \Lambda\pi^-,\, \Xi^- K^0,\, \phi\pi^-\Lambda ,\,
  K^+\pi^-\Xi^-,\, K^0K^-\Lambda ,\, K^-\pi^- p.
 % \phi K^0 \Xi^- ,\, K^0 K^0\Omega^- \,, 
  \nn
\end{align}
For bottom with $\bar b\to \bar c c\bar s$:
\begin{align} \label{Rlist1}
  & R_s^+ \to J/\Psi p ,\, \bar D^0 \Lambda_c ,\, D^- \Sigma_c^{++} ,\, 
  \pi^- \Delta^{++} ,\, J/\Psi \phi p, J/\Psi K^+\Lambda ,\, D_s^- D_s^+ p,\,
  D_s^- K^+ \Lambda_c , \\
  & \phantom{R_s^+ \to} 
      D_s^- K^0 \Sigma_c^{++} ,\, \bar D^0 D_s^+ \Lambda,\, 
      \bar D^0 \phi \Lambda_c ,\,
  \nn\\
  & R_s^0 \to K^0\Lambda,\, D^-\Lambda_c,\, \bar D^0 \Sigma_c^0 ,\, 
   \pi^- p ,\,   J/\Psi K^0\Lambda,\, D_s^- K^+\Sigma_c^0 ,\, 
   D_s^- K^0\Lambda_c ,\, \bar D^0 \phi \Sigma_c^0 , \nn\\
  & \phantom{R_s^0 \to }
  D^- D_s^+ \Lambda ,\, D^-\phi \Lambda_c \,, \nn\\
  & R_{ss}^0 \to \phi \Lambda ,\, J/\Psi \Lambda ,\, D_s^- \Lambda_c ,\,
   K^- p ,\, J/\Psi \phi\Lambda ,\, J/\Psi K^+ \Xi^- ,\, D_s^- D_s^+ \Lambda 
   ,\, D_s^- \phi \Lambda_c ,\,\bar  D^0 D_s^+ \Xi^-,  \nn
\end{align}
and with $\bar b\to \bar c u \bar d$:
\begin{align} \label{Rlist2}
  & R_s^+ \to D_s^- \Delta^{++},\, D_s^- \pi^+ p ,\, \bar D^0 \bar K^0 p,\, 
  \bar D^0 \pi^+ \Lambda ,\, D^- \bar K^0 \Delta^{++},\,
  \\
  & R_s^0 \to D_s^- p ,\, \bar D^0 \Lambda,\, D_s^- \pi^+ \Delta^0 ,\, 
  \bar D^0 \bar K^0 \Delta^0,\, D^- \bar K^0 p,\, D^-\pi^+\Lambda,\,
  \nn\\
  & R_{ss}^0 \to \bar D^0 \Xi^0 ,\, D_s^- \bar K^0 p,\, 
   \bar D^0 \bar K^0 \Lambda,\, \bar D^0 \pi^+ \Xi^-
   \,. \nn
\end{align}

\begin{figure}
\PSbox{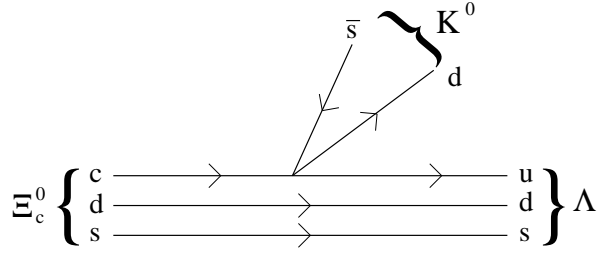 hoffset=50 voffset=20 hscale=50
vscale=50}{7.0in}{2.5in}
\caption[A Cabibbo-suppressed decay mode]{A Cabibbo-suppressed decay mode of the $\Xi_c^0$ that mimics the process in Figure \ref{fig:nonleptonic3}.}
\label{fig:mimic}
\end{figure}

In general, the quantum numbers of the final states in
eqs.~(\ref{Tlist}-\ref{Rlist2}) are not sufficient to tell us that the initial
state was exotic. Weak decays of $\Lambda_{b,c}$ and $\Xi_{b,c}$ can mimic some
of these channels through Cabibbo-suppressed or penguin
transitions, as shown in Figure \ref{fig:mimic}.\footnote{In one case there are even Cabibbo-allowed $\Lambda_b$ decays
  that mimic the $R_{ss}^0$ decaying through $\bar b\to \bar c c\bar s$.} 
Exceptions are the $T_s^-$ decays in Eq.~(\ref{Tlist}) and the $R_s^+$ decays
$(\bar b \to \bar c u\bar d)$ in Eq.~(\ref{Rlist2}): for these two cases
the final quantum numbers, $ddd$ and $\bar c s uuu$ respectively, are sufficiently exotic.  In
other cases, kinematic information is necessary. For two cases, the kinematic
information is fairly minimal: $T_{ss}^-$ ($\bar c\to \bar s u\bar d$) only has
contamination from weak $b$-baryon decays, and $R_s^+$ ($\bar b\to \bar c c\bar
s)$ only has contamination from weak $c$-baryon decays.

%%%%%%%%%%%%
%%%%%%%%%%%%%

The dynamics of nonleptonic decays is complicated, and some channels
may be suppressed relative to others. Therefore it is desirable to search in as
many channels as possible. If we consider measurements of multiple weak decays, there are isospin relations between the nonleptonic decays of the $R_s$
states.  Such relations can be derived using the method of \cite{savage}.  For example, for the process $\bar{b}\to \bar{c}d\bar{u}$, the weak Hamiltonian has the quantum numbers $(\bar{b}c)(\bar{d}u)$ and therefore has isospin 1.  It can be written as $H_{\alpha\beta}$, where the indices $\alpha$, $\beta$ are symmetric and can take the values 1, 2; the only nonzero component is $H_{11}$.  On the other hand, the weak Hamiltonian for the process $\bar{b}\to \bar{c}c\bar{s}$ is an isospin scalar.  We can combine these Hamiltonian with mesons and baryons to produce all possible scalar terms.  For example, to look at decays of the $R_a$ pentaquarks to a $K$ meson plus the isospin scalars $\Lambda$ and $J/\Psi$, we may write the effective Hamiltonian as
%%%%%%%%%%%%%%%%%%%%%
\begin{align}
a\bar{R}_\alpha H K^\alpha \Lambda J/\Psi,
\end{align} 
%%%%%%%%%%%%%%%%%%%%%
where $a$ is an unknown coefficient.  This tells us that the decay rates for $R_s^+\to k^+\Lambda J/\Psi$ and $R_s^0\to K^) \Lambda J/\Psi$ are both proportional to $|a|^2$, giving the first isospin relation in Eq. (\ref{riso}) below.  More complicated effective Hamiltonians can arise when the final state includes particles that are isospin doublets (such as $N$ or $K$) or triplets (such as $\Sigma$ or $\pi$).

Isospin relations for the $\bar b \to\bar c c\bar s$ transition include
%%%%%%%%%%%%%%%%%%%
\begin{align} \label{riso}
 \Gamma(R_s^+\to J/\Psi K^+\Lambda) & = \Gamma(R_s^0\to J/\Psi K^0 \Lambda) \,,
  \nn \\
 \Gamma(R_s^+\to \bar D^0 \phi\Lambda_c) & =\Gamma(R_s^0\to D^- \phi \Lambda_c) \,,
   \nn \\
 \Gamma(R_s^+\to D_s^- K^+ \Lambda_c) & = \Gamma(R_s^0\to D_s^- K^0 \Lambda_c)\,,\nn \\
\Gamma(R_s^+ \to \bar{D}^0\Lambda_c) & = \Gamma(R_S^0\to D^-\Lambda_c) \nn \\
2\Gamma(R_s^+\to \bar{D}^0\Sigma_c^+) & = \Gamma(R_s^+\to D^-\Sigma_c^{++}) \nn \\
& = \Gamma(R_s^0\to\bar{D}^0\Sigma_c^0) \nn \\
& = 2\Gamma(R_s^0\to D^-\Xi_c^+), \nn \\
\Gamma(R_{ss}^0\to\bar{D}^0\Xi_c^0) & = \Gamma(R_{ss}^0\to D^-\Xi_c^+), 
\end{align}
%%%%%%%%%%%%%%%
while for $\bar b \to \bar c u \bar d$,
%%%%%%%%%%%%%%%%%
\begin{align}
 \Gamma(R_{ss}\to D_s^- \Delta^{++} K^-) 
  & = 3\Gamma(R_{ss}\to D_s^- \Delta^+ \bar K^0) \,,\nn\\
 \Gamma(R_s^+\to \Delta^{++} D_s^+) 
 & = 3\Gamma(R_s^0 \to \Delta^+ D_s^-) \,.
\end{align}
%%%%%%%%%%%%%%%%%%%
For $\bar c\to \bar s d\bar u$ transitions of $T_{a}$, the isospin relations may be harder to test.  We find, for example, 
%%%%%%%%%%%%%%%%%
\begin{align}
  \Gamma(T_s^-\to \pi^-\Sigma^0K^0) & = \Gamma(T_s^-\to \pi^0\Sigma^-K^0) \nn \\
     \Gamma(T_{ss}^-\to \phi \pi^0 \Sigma^-)
     & =\Gamma(T_{ss}^-\to \phi \pi^-\Sigma^0) \,,\nn \\
   \Gamma(T_{ss}^-\to \phi \bar K^0 \Delta^-) 
    & = \Gamma(T_{ss}^-\to \phi  K^-\Delta^0)\,, \nn \\
  2\Gamma(T_{ss}^-\to  \bar K^0 \pi^0 \Delta^-)
    & =3 \Gamma(T_{ss}^-\to  \bar K^0 \pi^-\Delta^0) \,.
\end{align}
%%%%%%%%%%%%%%%%%%

\begin{figure}
\PSbox{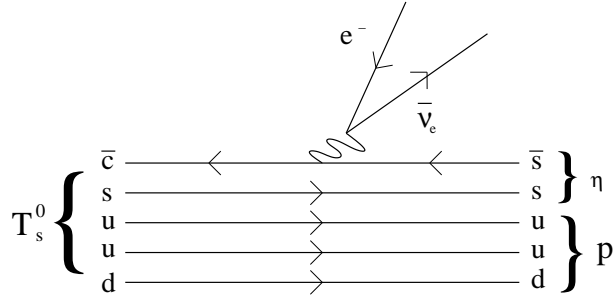 hoffset=50 voffset=20 hscale=50
vscale=50}{7.0in}{2.5in}
\caption[A semileptonic decay mode]{A semileptonic decay mode, $T_s^0\to \eta pe^-\bar{\nu}_e$.} 
\label{fig:semileptonic}
\end{figure}

The $T_a$ and $R_a$ states can also decay semileptonically, as in Figure \ref{fig:semileptonic},  and there are isospin and $SU(3)$ relations between semileptonic decays. 

%%%%%%%%%%%%
%%%%%%%%%%%%%

For
$T_a$ states decaying with a single baryon in the final state, the isospin
relation 
%%%%%%%%%%%%
\begin{align}
\Gamma(T_s^0 \rightarrow p e \bar \nu_e) =\Gamma(T_s^- \rightarrow n
e\bar \nu_e) 
\end{align}
%%%%%%%%%%%%%%%%%%%%%
relates the Cabibbo-allowed $T_s^0$ and $T_s^-$ decays.  For
decays with a baryon and a meson in the final state, we find the following
isospin relations:
%%%%%%%%%%%
\begin{align}
  2\Gamma(T_s^0 \rightarrow \pi^0 p e \bar \nu_e) 
  &=\Gamma(T_s^0 \rightarrow \pi^+ n e \bar \nu_e)
  =2\Gamma(T_s^- \rightarrow \pi^0 n e \bar \nu_e)
  =\Gamma(T_s^- \rightarrow \pi^- p e \bar \nu_e),
  \\
%\end{equation}
%\begin{equation}
  2\Gamma(T_s^0 \rightarrow K^+ \Sigma^0  e \bar \nu_e) 
  &=\Gamma(T_s^0 \rightarrow K^0 \Sigma^+ e \bar \nu_e)
  =\Gamma(T_s^- \rightarrow K^+ \Sigma^- e \bar \nu_e)
  =2\Gamma(T_s^- \rightarrow K^0 \Sigma^0  e \bar \nu_e),
  \nn\\
%\end{equation}
%and
%\begin{equation}
 \Gamma(T_s^0 \rightarrow \eta p  e \bar \nu_e)
  &=\Gamma(T_s^- \rightarrow \eta n  e \bar \nu_e), \nn \\
 \Gamma(T_s^0 \rightarrow K^+ \Lambda  e \bar \nu_e)
 &  =\Gamma(T_s^- \rightarrow K^0 \Lambda  e \bar \nu_e).\nn \hspace{1cm}
\end{align}
%%%%%%%%%%%%
For semileptonic decays involving $\bar c\to \bar s$,
the weak Hamiltonian transforms as a ${\bf 3}$ of $SU(3)$.   To find $SU(3)$ relations, we will need the baryon octet $B^a_b$ and the pseudoscalar meson octet $M^a_b$, where the matrix of octet baryons is written as
%%%%%%%%%%%%
\begin{align}
B =\left(\begin{array}{ccc} \frac{\Sigma^0}{\sqrt{2}}+\frac{\Lambda}{\sqrt{6}} & \Sigma^+ & p \\
\Sigma^-  & -\frac{\Sigma^0}{\sqrt{2}}+\frac{\Lambda}{\sqrt{6}} & n \\
\Xi^- & \Xi^0 & -\sqrt{\frac{2}{3}}\Lambda \end{array}\right), 
\end{align}
%%%%%%%%%%%
and the matrix of pseudoscalar octet mesons is
%%%%%%%%%%
\begin{align}
M=\left(\begin{array}{ccc} \frac{\pi^0}{\sqrt{2}}+\frac{\eta}{\sqrt{6}} & \pi^+ & K^+ \\
\pi^-  & -\frac{\pi^0}{\sqrt{2}}+\frac{\eta}{\sqrt{6}} & K^0 \\
K^- & \bar{K}^0 & -\sqrt{\frac{2}{3}}\eta \end{array}\right).
\end{align}
%%%%%%%%%%%
We can contract $T_a$ and the Hamiltonian $H^a$ (whose only nonzero component is $H^3$) with $B^a_b$ and $M^a_b$ to obtain all possible scalar terms; as in the isospin case, this will tell us how the decay rates are related.  For example, with decays to a single baryon plus leptons, we get
%%%%%%%%
\begin{align}
H_{eff} = cT_aH^bB^a_b, 
\end{align}
%%%%%%%%
and we find
%%%%%%%%%%%
\begin{equation}
  2\Gamma(T_s^0 \rightarrow p e \bar \nu_e) = 2\Gamma(T_s^-\rightarrow n e \bar \nu_e)
    = 3\Gamma(T_{ss}^- \rightarrow \Lambda e \bar \nu_e).
\end{equation}
%%%%%%%%%%%
There is also one (independent) $SU(3)$ relation between the semileptonic decays
to a meson and a baryon for the strangeness -1 states:
%%%%%%%%%%
\begin{equation} \label{su3}
  2\Gamma(T_s^0 \rightarrow \eta p e \bar \nu_e)
   -2\Gamma(T_s^0 \rightarrow K^+ \Lambda e \bar \nu_e)
  = \Gamma(T_s^0 \rightarrow K^0 \Sigma^+ e \bar \nu_e)
   -\Gamma(T_s^0 \rightarrow \pi^+ n e \bar \nu_e).
\end{equation}
%%%%%%%%%%%%%%%
Similar results can be derived for semileptonic $R_a$ decays.  Isospin relations include,
%%%%%%%%%%%%%%%
\begin{align}
\Gamma(R_s^+\to\bar{D}^0\Lambda e^+\nu_e) & = \Gamma(R_s^0\to D^-\Lambda e^+\nu_e) \nn \\
\Gamma(R_s^+\to D_s^- pe^+\nu_e) & = \Gamma(R_s^0\to D_s^- n e^+\nu_e) \nn \\
\Gamma(R_{ss}^0\to\bar{D}^0\Xi^-e^+\nu_e) & = \Gamma(R_{ss}^0\to D^-\Xi^0e^+\nu_e).
\end{align}
%%%%%%%%%%%%%%%%%%%%%
The weak Hamiltonian for these $\bar{b}\to\bar{c}$ decays is an $SU(3)$ singlet, and so it is not possible to derive an $SU(3)$ relation analogous to Eq. (\ref{su3}).  

A crucial aspect of the detectability of these exotic states is their production
rate via fragmentation of the heavy quark.  A crude estimate,
inspired by the fact that $\Lambda_b$ production via fragmentation is a factor of $\sim 0.3$ less than $B$ production, is that every additional quark (or
antiquark) costs 0.3. This suggests that the production rate for
the $R_s$ (or $T_s$) states may be $\sim 10^{-2}$ of the $B_s$ (or $D_s$) mesons, in agreement with earlier estimates for pentaquark production \cite{moinester}.  

\subsection{Other Possible $S$-wave States}

The combinations of quarks $\bar Q suud$ and $\bar Q sudd$ do not have to have
$I=1/2$. For example, an $S$-wave $(ud)$ quark pair in a spin one configuration can
be a color antitriplet if it is in an $I=1$ configuration. Combining this with
the other $u$ quark gives the possibility of an $I=3/2$ final state. However, it
appears likely that these states are heavier than the states we have been
considering. Phenomenological evidence for this comes from the fact that the
$\Sigma_c$ is heavier than the $\Lambda_c$ (and the $\Sigma$ is heavier than the
$\Lambda$). These other isospin states will decay to the ones we are considering
via emission of a photon, and, if the mass splitting is large enough, by emission
of a pion.

\begin{figure}
\PSbox{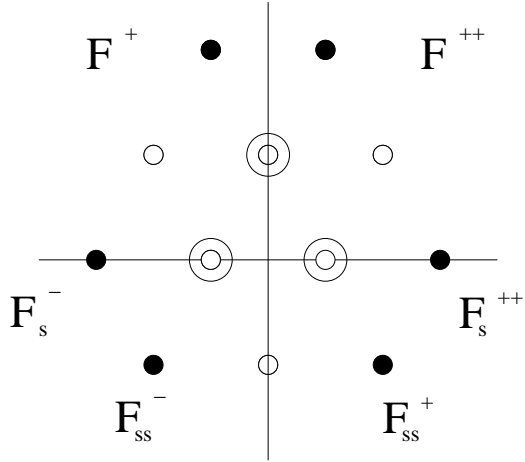 hoffset=20 voffset=20 hscale=80
vscale=80}{3.0in}{3.0in}
\caption[The ${\bf \overline {15}}$ representation]{The ${\bf \overline {15}}$ representation of $SU(3)$.  The exotic states are shown in black, cryptoexotic states in white.}
\label{fig:fifteenbar}
\end{figure}

It is also possible to construct $S$-wave pentaquark states in which a heavy
quark, say a charm, is part of one of the diquarks.  The exotic states of
this type are part of the ${\bf \overline {15}}$ representation of $SU(3)$ (Figure \ref{fig:fifteenbar}).

%%%%%%%%%%%%
%%%%%%%%%%%%%%

 They include,
%%%%%%%%%%%%%%%%%%%%
\begin{align}
  F_2^{11} &= F_s^- = \bar u (ds)(cd)  \,,\qquad
  F_1^{22} = F_s^{++} = \bar d (su)(cu) \,, \qquad 
  F_3^{11} = F_{ss}^- = \bar u (ds)(cs) \,,\nn\\ 
  F_3^{22} &= F_{ss}^+ = \bar d (su)(cs) \,,\qquad
  F_1^{33} = F^{++} = \bar s (ud)(cu)   \,,\qquad
  F_2^{33} = F^+ = \bar s (ud)(cd) \,.
\end{align}
%%%%%%%%%%%%%%%%%%%%%
(The other 11 members of this multiplet are cryptoexotic states.) The charge-two states are particularly distinctive; if they are stable against
strong decay, they could be detected, for example, via the mode
$F^{++} \rightarrow p \pi^+$.  Unfortunately, we are not able to draw conclusions about the masses of the states in the ${\bf \overline{15}}$  multiplet from the observed $\Theta$ mass.  They may be heavy enough to decay strongly to a $B$ or $D$ meson plus a baryon.

$F^{++}$ and $F^+$ form an isospin doublet, while $F_s^-$, $F_s^{++}$ have $I=\frac{3}{2}$, and $F_{ss}^-$, $F_{ss}^+$ have $I=1$ (their isospin partners being cryptoexotic states). If we assume these states decay weakly, there are a number of isospin relations for the decays of the charmed $F$ states:
%%%%%%%%%%%%%%%%%%%
\begin{align}
 \Gamma(F_s^- \to \pi^- \Xi^- e^+ \nu) &=
 \Gamma(F_s^{++} \to\pi^+\Xi^0 e^+\nu) \,, \\
 \Gamma(F_s^- \to K^- \Sigma^- e^+ \nu) &=
  \Gamma(F_s^{++} \to \bar K^0 \Sigma^+ e^+ \nu) \,,\nn\\
 \Gamma(F_{ss}^- \to K^- \Xi^- e^+ \nu) &= 
  \Gamma(F_{ss}^+ \to \bar K^0 \Xi^0 e^+\nu) \,,\nn\\
\Gamma(F^{++} \to\eta p e^+\nu) &=
   \Gamma(F^+ \to \eta n e^+ \nu) \,,\nn\\
 \Gamma(F^{++} \to K^+\Lambda e^+\nu) &=
  \Gamma(F^+ \to K^0 \Lambda e^+\nu) \,,\nn\\
 2\Gamma(F^{++} \to \pi^0 p e^+\nu) &= 
  \Gamma(F^{++} \to \pi^+ n e^+ \nu) = \Gamma(F^+ \to \pi^- p e^+ \nu) 
   = 2\Gamma(F^+ \to \pi^0 n e^+ \nu) \,,\nn\\
 2\Gamma(F^{++} \to K^+ \Sigma^0 e^+ \nu) &= 
 \Gamma(F^{++} \to K^0 \Sigma^+ e^+ \nu) = \Gamma(F^+ \to K^0 \Sigma^0 e^+ \nu) 
 = 2\Gamma(F^+ \to K^+ \Sigma^- e^+ \nu) \,. \nn
\end{align}
%%%%%%%%%%%%%%%%%%%%
The analogous states with a bottom quark in one of the diquarks are
%%%%%%%%%%%%%%%%%%%%
\begin{align}
  F_{b2}^{11} &= F_{bs}^{--} = \bar u (ds)(bd)  \,,\qquad
  F_{b1}^{22} = F_{bs}^{+} = \bar d (su)(bu) \,, \qquad 
  F_{b3}^{11} = F_{bss}^{--} = \bar u (ds)(bs) \,,\nn\\ 
  F_{b3}^{22} &= F_{bss}^0 = \bar d (su)(bs) \,,\qquad
  F_{b1}^{33} = F_b^{+} = \bar s (ud)(bu)   \,,\qquad
  F_{b2}^{33} = F_b^0 = \bar s (ud)(bd) \,.
\end{align}
%%%%%%%%%%%%%%%%%%%%%
Again, some have a distinctive charge, -2 in this case.  

There are isospin relations for these states as well, including semileptonic decays,
%%%%%%%%%%%%%%%
\begin{align}
\Gamma(F_{bs}^{--}\to D^0\Sigma^-e^-\bar{\nu}_e) & = \Gamma(F_{bs}^+\to D^+\Sigma^+e^-\bar{nu}_e) \nn \\
\Gamma(F_{bss}^{--}\to D^0\Xi^-e^-\bar{\nu}_e) & = \Gamma(F_{bss}^0\to D^+\Xi^0e^-\bar{\nu}_e) \nn \\
\Gamma(F_b^+\to K^+\Lambda_c e^-\bar{\nu}_e) & = \Gamma(F_b^0\to K^0\Lambda_c e^-\bar{\nu}_e) \nn\\
\Gamma(F_b^+\to D_s^+pe^-\bar{\nu}_e) & = \Gamma(F_b^0\to D_s^+n e^-\bar{\nu}_e), 
\end{align}
%%%%%%%%%%%%%%%%%%%%%%
and nonleptonic decays,
%%%%%%%%
\begin{align}
\Gamma(F_b^+\to p D_s^+ D_s^-) & = \Gamma(F_b^0\to n D_s^+ D_s^-) \nn \\
\Gamma(F_b^+\to p\Lambda J/\Psi) & = \Gamma(F_b^0\to n\Lambda J/\Psi) \nn \\
\Gamma(F_b^+\to K^+\Lambda J/\Psi) & = \Gamma(F_b^0\to K^0\Lambda J/\Psi) \nn \\
\Gamma(F_b^+\to \eta \bar{D}^0\Lambda_c) & = \Gamma(F_b^0\to \eta D^- \Lambda_c) \nn \\
\Gamma(F_{bs}^+\to \Sigma^+D^+D_s^-) & = \Gamma(F_{bs}^{--}\to \Sigma^-D^0D_s^-) \nn \\
\Gamma(F_{bs}^+\to K^0 \Sigma^+ J/\Psi) & = \Gamma(F_{bs}^{--}\to K^-\Sigma^-J/\Psi) \nn \\
\Gamma(F_{bss}^0\to \Xi^0D^+D_s^-) & = \Gamma(F_{bss}^{--}\to\Xi^-D^0D_s^-) \nn \\
\Gamma(F_{bss}^0\to \pi^+J/\Psi \Omega^-) & = \Gamma(F_{bss}^{--}(\pi^-J/\Psi \Omega^-). 
\end{align}
%%%%%%%%%%%
Because we have no estimate for the masses of these states, it is uncertain whether they will ever be seen experimentally.  If they are heavy enough to decay strongly, they may appear as broad resonances that would never be detectable.  On the other hand, the $\Theta^+$ itself decays strongly, and yet it is narrow enough to produce a clear experimental signal.  The structure and dynamics of the $F$ and $F_b$ states could conceivably be such that they, too, will be narrow enough to be visible, whether they decay strongly or weakly.

\section{Summary and Discussion}

In this chapter, we have explored some of the implications of the diquark model for heavy exotic baryons.  Most notably, we have shown that some such states might be relatively accessible to an experimental search.  An observation of heavy pentaquarks of the form $T_a$ or $R_a$ that were stable against strong decay, in addition to being exciting in its own right,  would provide support for the diquark interpretation of the $\Theta^+$.  While the predictions here are quite rough, they suggest that  an experimental search for such states could be worthwhile.  The decay channels and isospin relations listed here could serve as guidelines for what to look for in such an experiment.

It may also be useful to study these pentaquarks further in lattice QCD, to help determine whether our estimates are reasonable.  Lattice calculations are the only currently available method of working directly with QCD, as opposed to using a particular constituent quark model or a general symmetry-based method like the large $N_c$ expansion.  As noted in the Introduction, lattice studies of the $\Theta^+$ have suffered from a number of problems and have generally been inconclusive; however, computing power and methodology are continually improving, and it is likely that these problems will be fixed in due time.  It would be quite interesting to see whether a lattice calculation would agree with our estimates for the $T_a$ and $R_a$ masses. 
  
\chapter{Heavy Pentaquarks in the Large $N_c$ Expansion}

\section{The Large $N_c$ Formalism}

As mentioned in Chapter 1, the strong coupling constant $\alpha_s$ is not a useful expansion parameter at energies of $\sim \Lambda_{QCD}$ or below, because $\alpha_s >1$ in this regime.  However, 't Hooft  showed  in the 1970's  \cite{thooft} that there is another expansion parameter valid in any regime of QCD: $1/N_c$, where $N_c$ is the number of colors ({\it i.e.}, the gauge group is $SU(N_c)$).  't~Hooft showed that a Feynman diagram for vacuum quark-gluon interactions is proportional to 
%%%%%%%%%%%%%%
\begin{align} \label{qg}
(g^2N_c)^{\frac{1}{2}V_3+V_4}N_c^\chi, 
\end{align}
%%%%%%%%%%%%%% 
where $g$ is the quark-gluon coupling constant, $V_n$ is the number of $n$-point vertices in the diagram, and $\chi$ is the diagram's Euler characteristic, a topological invariant defined by
%%%%%%%%%%%
\begin{align}
\chi = 2-2H-L,
\end{align}
%%%%%%%%%%%%%%
with $H$ being the number of handles (formed by nonplanar gluon exchange), and $L$ being the number of quark loops.  

It is evident from Eq. \eqref{qg} that Feynman diagrams with increasing numbers of vertices grow with increasingly large powers of $N_c$, unless the $N_c\to\infty$ limit is taken with $g^2N_c$ held fixed.  This procedure is known as the 't Hooft limit, and can be implemented by rescaling $g\to g/\sqrt{N_c}$.  After this rescaling, Eq. \eqref{qg} tells us that diagrams with nonplanar gluon exchange will be suppressed by $1/N_c^2$ for each such gluon, while diagrams with quark loops will be suppressed by $1/N_c$ for each loop. 

At first sight, this appears to be a very strange sort of expansion.  After all, in nature, $N_c$ is fixed at 3.  For large $N_c$, baryons become unwieldy-looking objects made of $N_c$ quarks.   However, the $1/N_c$ expansion turns out to be a useful way of quantifying spin-flavor symmetry breaking effects.  Spin-flavor symmetry for baryons formally becomes an exact symmetry in the 't Hooft limit; this symmetry is explicitly broken by corrections suppressed by powers of $1/N_c$.   In the expansion, $N_c$ is always taken to be an odd integer; an even $N_c$ would produce baryons that were bosons rather than fermions.  Putting in the real-world value $N_c=3$ gives an expansion parameter of $1/3$; thus $1/N_c$ effects are suppressed by about the same amount as flavor $SU(3)$ breaking effects.  

Large $N_c$ is a rich and complicated subject that has been treated only briefly here.  For a recent review, see \cite{manoharnc}.   See \cite{march} for a somewhat different approach to $1/N_c$ calculations than that employed here.  

Now, let us consider more closely the large $N_c$ treatment of baryons.  In the $N_c \rightarrow \infty$ limit, baryons form irreducible representations of contracted spin-flavor $SU(6)_c$; for finite $N_c$, this symmetry is broken, generating mass splittings within each representation.  The symmetry breaking can be parameterized using polynomials in the $SU(6)$ generators:
%%%%%%%%%%%%%%%%%%%%%%%%%%%%%%%%%%%%%%%%%
\begin{align} \label{generators}
  S^i & \equiv  q^\dag (\frac{\sigma^i} {2} \otimes \openone)q \nn \\
 T^a & \equiv  q^\dag (\openone \otimes \frac{\lambda^a}{2})q \nn \\
  G^{ia} & \equiv  q^\dag  (\frac{\sigma^i}{2} \otimes \frac{\lambda^a}{2})q,
\end{align}
%%%%%%%%%%%%%%%%%%%%%%%%%%%%%%%%%%%%%%%%%
where $q^\dag$ and $q$ are quark creation and annihilation operators, $\sigma^i$ are the Pauli matrices, and $\lambda^a$ are the Gell-Mann matrices.  An $n$-body operator, which acts on $n$ quark lines in a baryon, comes with a factor $N_c^{1-n}$.  The generator $G^{ia}$ sums coherently over all the quark lines and hence is order $N_c$.  $T^a$ may also sum coherently when three or more flavors are considered.  (When the discussion is limited to two flavors, as in \cite{carlson99}, the isospin is fixed in the large $N_c$ limit, so $T^a$ is order 1.) Thus a given $n$-body operator contributes at order $N_c^{1-n-m-p}$, where $m$ is the number of times $G^{ia}$ appears and $p$ is the number of times $T^a$ appears.  To describe mass splittings, one constructs all possible scalar operators up to a given order in $1/N_c$; each such operator appears in the expansion with an unknown coefficient of order unity.  Depending on the symmetry of the baryon states under consideration, there may be operator reduction rules allowing some operators to be eliminated; for example, for states that are completely spin-flavor symmetric, it is unnecessary to include operators in which flavor indices are contracted using $\delta_{ab}$ or $\epsilon_{abc}$.  A complete list of the reduction rules for completely symmetric states is given in \cite{dashen}. 

It is important to note that the irreducible representations of $SU(6)_c$ are not identical to those of the uncontracted $SU(6)$.  In the contracted symmetry group, a single $SU(6)$ representation can split into several ``towers" whose masses are not the same even in the $N_c\to \infty$ limit \cite{pirjol, pirjol2}.  In the two-flavor case, when the symmetry group is $SU(4)_c$, such a tower is labeled by $K=0,1/2, 1, \dots$, where the states in the tower have spin and isospin satisfying $|I-J| \leq K$.  Baryons with strangeness form separate towers, labeled by $K=\frac{1}{2}n_s$, where $n_s$ is the number of strange quarks.

\subsection{Pentaquarks in Large $N_c$}

\begin{figure}[h]
\PSbox{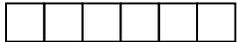 hoffset=40 voffset=10 hscale=80
vscale=80}{6.8in}{1.0in}
\caption[The ${\bf 28}$ of $SU(3)$]{The ${\bf 28}$ of $SU(3)$.  This representation has $r=1$ but exoticness $E=2$.}
\label{fig:twentyeight}
\end{figure}

The $1/N_c$ expansion has recently been extended to exotic baryons, including partners of the $\Theta$ \cite{cohen, manohar} and heavy pentaquarks in which the antiquark is a $\bar c$ or a $\bar b$  \cite{manohar}.  In order to treat exotic baryons in a group-theoretical way, the concept of ``exoticness'' is defined \cite{jm} as the minimum number of $q\bar{q}$ pairs needed to construct a baryon with a given set of quantum numbers.  The precise definition of exoticness requires some care.  For an $SU(3)$ representation with Dynkin indices $(p,q)$, the corresponding Young tableau has $N_c+3r$ boxes, where $N_c+3r=p+2q$ and $p+q \geq r \geq 0$.  The integer $r$ appears to correlate with the number of additional quarks beyond $N_c$, and in an earlier work, \cite{petrov}, $r$ was called ``exoticness.''  However, this is not quite what one would like ``exoticness'' to mean.  For example, the ${\bf 28}$ of $SU(3)$ has a Young tableau with six boxes (Figure \ref{fig:twentyeight}), and thus has $r=3$.  But it is not possible to construct a state in this representation with four quarks and one antiquark, as one can see in a straightforward manner by calculating the direct product ${\bf \overline{3}}\otimes{\bf 3}\otimes{\bf 3}\otimes{\bf 3}\otimes{\bf 3}$ in $SU(3)$ and noting that ${\bf 28}$ does not appear in the result.  The proper way to define exoticness for the representation $(p,q)$ is
%%%%%%%%%%%%%
\begin{align} \label{ex}
E = \left\{\begin{array}{lc} r & {\rm if} \;  r \leq q, \\ 2r-q & {\rm if} \;  r \geq q.\end{array}\right.
\end{align}
%%%%%%%%%%%%%%
This gives the intuitive result for $E$: a baryon with exoticness $E$ contains $N_c+E$ quarks and $E$ antiquarks.  (A cryptoexotic pentaquark has $E=0$; $r$ is defined using the {\it minimal} diagram that yields a state with a given set of quantum numbers.)  For the ${\bf 28}$, $(p,q)=(6,0)$; $r=1>q$, so Eq. (\ref{ex}) gives $E=2$.

\begin{figure}[h]
\PSbox{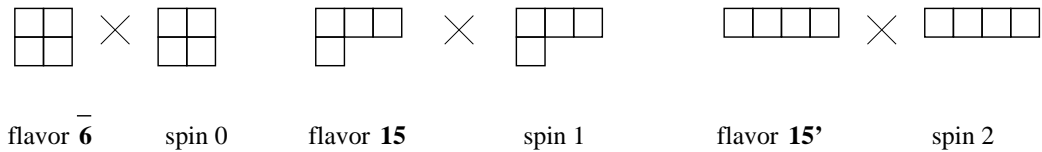 hoffset=10 voffset=10 hscale=60
vscale=60}{6.8in}{1.7in}
\caption[Spin-flavor decomposition of the ${\bf 126}$ of $SU(6)$]{Young tableaux showing the flavor $SU(3)$ $\times$ spin $SU(2)$ decomposition of the completely symmetric ${\bf 126}$ of $SU(6)$, giving three different multiplets.  States in the completely symmetric representation are distinctive because the Young tableaux for their spin and flavor look the same.}
\label{fig:spinflavor2}
\end{figure}

In \cite{manohar}, Jenkins and Manohar derive relations between the masses and decay rates of various multiplets of $E=1$ baryons, both light and heavy.  Their work assumes that the pentaquark states are in the completely symmetric representation of spin-flavor $SU(6)$ and thus have positive parity.  In the case of heavy pentaquarks, this means that the four light quarks are in the ${\bf 126}$ of $SU(6)$, which decomposes into three different spin-flavor states: ${\bf \overline{6}}_0$, ${\bf 15}_1$, and ${\bf 15^\prime}_2$ (Figure \ref{fig:spinflavor2}).

%%%%%%%%%%%%%%%%%%%%%%%%%%%%%%
%%%%%%%%%%%%%%%%%%%%%%%%%%%%%%%

The assumption of positive parity is sensible because it has been shown, in the context of a constituent quark model \cite{glozman}, that the hyperfine flavor-spin interactions between quarks in a hadron are most attractive for completely symmetric states.  However, as noted in the previous chapter,  a positive-parity pentaquark would need to have one quark in an orbitally excited $\ell = 1$ state in order to satisfy Fermi statistics.  It is not clear whether the resulting $P$-wave energy would always be sufficiently offset by the attractive flavor-spin interactions to make the positive-parity pentaquarks lighter than their negative-parity counterparts.  Thus it is worthwhile to apply the large $N_c$ formalism to the negative-parity case; this has been done in two recent papers.  One \cite{wessling}, by the present author, focused on heavy pentaquarks; the other \cite{pirjolschat} also looked at light pentaquarks and considered the possibility that both the $\Theta^+$ and the $\Theta_c$ seen by H1 may have negative parity.

\subsection{Non-Exotic Excited Baryons}

\begin{figure}[h]
\PSbox{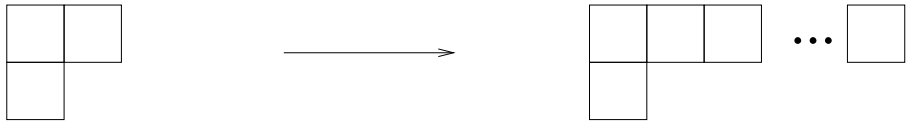 hoffset=40 voffset=10 hscale=80
vscale=80}{6.8in}{1.5in}
\caption[Extension of the excited baryons to large $N_c$]{Extension of the excited baryons to large $N_c$.  The top row of the tableau on the right has $N_c - 1$ boxes.}
\label{fig:excited}
\end{figure}

The $1/N_c$ expansion for excited baryons provides a model for working with states of mixed spin-flavor symmetry.  References \cite {carlson99,goity,pirjol, pirjol2} study excited baryons in the {\bf 70} of $SU(6)$; they generalize this representation for $N_c > 3$ as shown in the Young diagram in Figure \ref{fig:excited}. 

%%%%%%%%%%%%%%%%%%%%%%%%%%%%%%%%%%%%
In this picture, a baryon contains one excited quark, with angular momentum $\ell = 1$, and $N_c - 1$ ``core'' quarks, which are completely symmetric in spin-flavor $SU(6)$.  The expansion is made using two sets of $SU(6)$ generators: $s^i$, $t^a$, $g^{ia}$, acting on the excited quark; and $S^i_c$, $T^a_c$, $G^{ia}_c$, acting on the core.  The reduction rules for these operators are determined in \cite {carlson99}; the reduction rules for completely symmetric states apply to the core and excited operators separately, while additional rules govern the combining of the two types.   

\section{Negative-Parity Pentaquarks in Large $N_c$}

\begin{figure}[h]
\PSbox{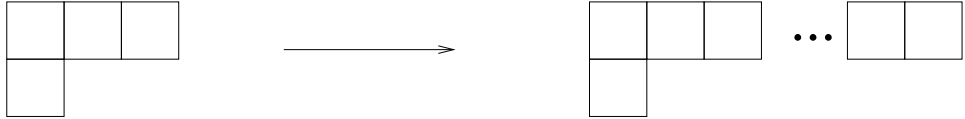 hoffset=40 voffset=10 hscale=80
vscale=80}{6.8in}{1.5in}
\caption[Extension of the negative-parity pentaquarks to large $N_c$]{Extension of the negative-parity pentaquarks to large $N_c$.  The top row of the tableau on the right now has $N_c$ boxes.}
\label{fig:negparitync}
\end{figure}

We wish to examine exotic negative-parity baryons containing $N_c + 1$ light quarks and one heavy antiquark, which can be extended to large $N_c$ in a similar manner to the excited baryons, as shown in Figure \ref{fig:negparitync}.  
%%%%%%%%%%%%%%%%%%%%%%%%%%%%%%
%%%%%%%%%%%%%%%%%%%%%%%%%%%%%%%
Here too, it makes sense to construct operators from two different sets of generators.  The term ``excited'' does not apply in this case, because the pentaquarks have no orbital angular momentum.  However, dividing the states into $N_c$ symmetrized quarks (which we will continue to call the ``core''), plus one extra, still captures their symmetry properties in a useful way.   The same operators and reduction rules constructed for the excited baryons in \cite{carlson99,goity,pirjol} may be used to describe the negative-parity pentaquarks.  In fact, the situation simplifies significantly in the pentaquark case, because the seven operators depending on $\ell$ all vanish.  We are left, at order $1/N_c$, with six linearly independent operators:\footnote{The expansion should also contain operators depending on the exoticness $E$, such as $E \openone$ and $\frac{1}{N_c} E^2$.  However, since $E=1$ for all the pentaquark states, these operators do not tell us anything new about the mass splittings.}
\begin{align}
%%%%%%%%%%%%%%%%%
O_1 & \equiv  N_c {\bf 1}, \nn \\ 
O_2 & \equiv  \frac{1}{N_c} S_c^2, \nn \\ 
O_3 & \equiv  \frac{1}{N_c} s^iS_c^i, \nn \\  
O_4 & \equiv  \frac{1}{N_c} t^aT_c^a, \nn \\
O_5 & \equiv  \frac{2}{N_c^2} t^a \{ S_c^i,G_c^{ia} \}, \nn \\
O_6 & \equiv  \frac{1}{N_c^2}g^{ia}S_c^iT_c^a.
\end{align}
%%%%%%%%%%%%%%%%%%
The {\bf 210} of $SU(6)$, which describes the $N_c + 1$ light quarks,  can be decomposed into flavor $\otimes$ spin to give seven different multiplets\footnote{The superscript $P$ indicates pentaquark states. Later we will use $E$ to denote excited baryons and $N$ for normal baryons.} (see, e.g., \cite{itzykson}): ${\bf 15}^{\prime P} _1$, ${\bf 15}_2^P$,  ${\bf 15}_1^P$, ${\bf 15}_0^P$,  ${\bf \bar 6}_1^P$,  ${\bf 3}_1^P$, and  ${\bf 3}_0^P$.  

%%%%%%%%%%%%%%%%%%%%%%%%%%%%%%
\begin{figure}[h]
\PSbox{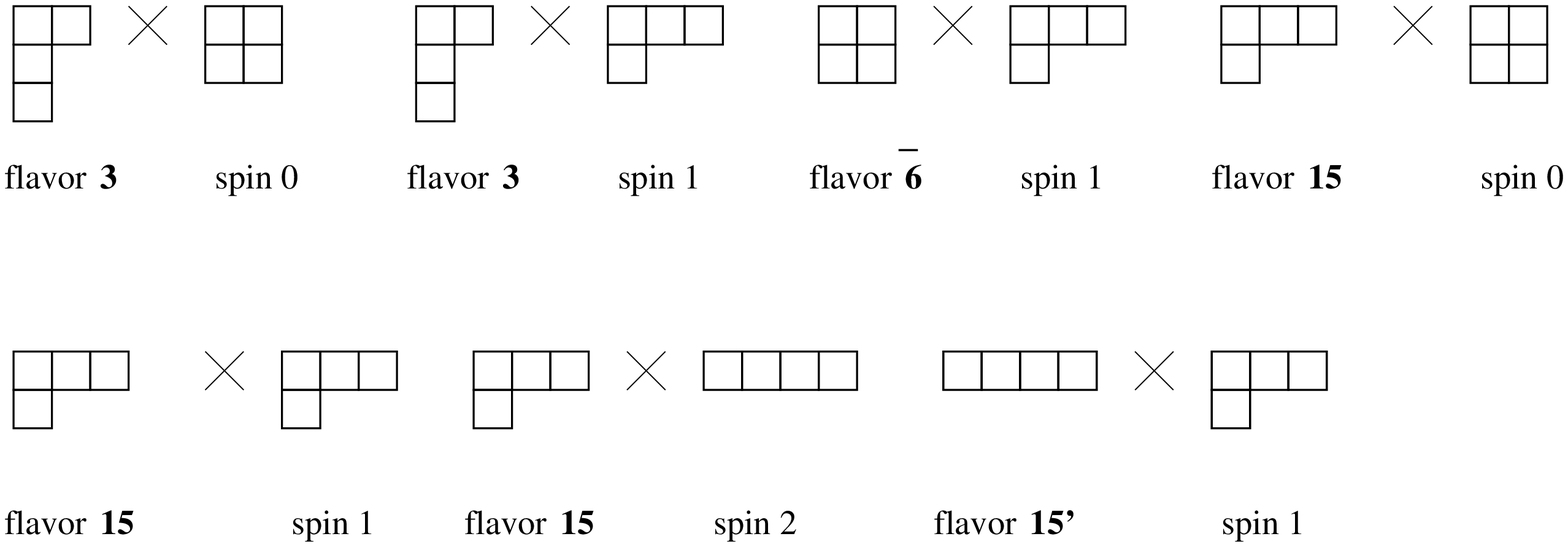 hoffset=10 voffset=10 hscale=60
vscale=60}{6.8in}{3.0in}
\caption[Spin-flavor decomposition of the ${\bf 210}$ of $SU(6)$]{Young tableaux showing the flavor $SU(3)$ times spin $SU(2)$ decomposition of the mixed-symmetry ${\bf 210}$ of $SU(6)$, giving seven different multiplets.}
\label{fig:spinflavor}
\end{figure}
%%%%%%%%%%%%%%%%%%%%%%%%%%%%%%%

In extending this decomposition to large $N_c$, the spin and strangeness of each state remain fixed; we add to the core a spin 0, isospin 0 combination of up and down quarks.  The flavor representations change with $N_c$; however, we will always use the $N_c = 3$ values for notational purposes in this paper. (We will also continue to use the term ``pentaquark'' for the large-$N_c$ analogues of such states.)  Figure \ref{fig:su3} contains diagrams of the four different flavor representations for $N_c=3$, and Figure \ref{fig:extension} illustrates how they are extended to larger values of $N_c$.

The ${\bf 3}_0^P$ is the multiplet called $T_a$ or $R_a$ in \cite{stewart} and the previous chapter; the three states have the flavor content $\bar Q uuds$, $\bar Q udds$, and $\bar Q udss$.  Interestingly, the large-$N_c$ version of this state can still be thought of in terms of the diquark model.  In this view, a ``diquark'' for arbitrary $N_c$ still consists of two quarks combined antisymmetrically in color and flavor; it may be written $\phi^a_{[\alpha \beta]}$, where $a$ is a flavor index and $\alpha$, $\beta$ are antisymmetrized color indices.  The full state is then
%%%%%%%%%%%%%%%%%%%%%%%%%%%%%%%%
\begin{equation}
T_a^{d_1...d_{(N_c-3)/2}} = \delta^\alpha_\beta \epsilon_{abc} \epsilon^{\gamma_1\gamma_2...\gamma_{N_c}} \bar Q^\alpha \phi^b_{\beta \gamma_1} \phi^c_{\gamma_2 \gamma_3} \phi^{d_1}_{\gamma_4 \gamma_5}...\phi^{d_{(N_c-3)/2}}_{\gamma_{N_c-1} \gamma_{N_c}}.
\end{equation}
%%%%%%%%%%%%%%%%%%%%%%%%%%%%%%%
The ${\bf 3}_0^P$ is the only one of the seven multiplets that can be constructed using Jaffe and Wilczek's original spin 0, flavor ${\bf \bar 3}$ diquarks.  However, if we also allow spin 1, flavor ${\bf 6}$ diquarks--called tensor diquarks  in \cite{shuryak} and ``bad'' diquarks in \cite{jaffexotica}--many other multiplets become possible.  The remainder of this chapter makes no reference to the diquark model; the results are model-independent.\footnote{In fact, there is a rather subtle issue regarding model-independence: do the results of this analysis follow from large-$N_c$ QCD alone, or do they depend on the large-$N_c$ quark model?  The ordinary ground-state baryons are stable in the large-$N_c$ limit, with properties entirely determined by symmetry; the quark model in this case is just a convenient way of counting states, introducing no dynamical assumptions beyond large-$N_c$ QCD.  However, this is not true for the excited baryons; they have widths that go as $N_c^0$, and so using the quark model for them does introduce new dynamical assumptions.  References \cite{cohen2} address this problem in detail, and show that the treatment of excited baryons in large $N_c$ makes sense. A recent paper by Cohen and Lebed \cite{cohen05} extends this treatment to exotic multiplets such as the large $N_c$ analogue of the ${\bf \overline{10}}$, which contains the $\Theta^+$.}

It is straightforward to determine what the ``core'' of each state should look like: $S_c = S_{total} \pm \frac{1}{2}$, and the core flavor representation can be written in Dynkin index notation as $(p, q) = (2S_c, \frac{N_c-2S_c}{2})$.  For six of the seven multiplets, there is only one possible value of $S_c$.  In particular: $S_c = \frac{1}{2}$ for the ${\bf 15}_0^P$,  ${\bf \bar 6}_1^P$,  ${\bf 3}_1^P$, and  ${\bf 3}_0^P$ states;  $S_c = \frac{3}{2}$ for the ${\bf 15}^{\prime P} _1$ and ${\bf 15}_2^P$ states.  The ${\bf 15}_1^P$ state is somewhat more complicated, because the flavor-spin decomposition of the totally symmetric representation of $SU(6)$ also contains a ${\bf 15}_1$.  The correct core is a linear combination of $S_c = \frac{3}{2}$ and  $S_c = \frac{1}{2}$, whose coefficients can be determined using Casimir operators.  Reference \cite{carlson99} finds the analogous coefficients for total spin $S$ and a core of $N_c - 1$ quarks; we may simply use their result with $S = 1$ and $N_c \rightarrow N_c + 1$: 
%%%%%%%%
\begin{equation} \label{coef}
  \sqrt{\frac{N_c+5}{3(N_c+1)}} \left | S_c = \frac{3}{2} \right > - \sqrt{\frac{2(N_c-1)}{3(N_c+1)}} \left | S_c = \frac{1}{2} \right >.
\end{equation}
%%%%%%%%
Evaluating the matrix elements of some of these operators, particularly $O_5$, is rather a nontrivial task; it cannot be done by a simple $N_c \rightarrow N_c + 1$ substitution.  One method of evaluation is to construct the wavefunction for each state as in section II of \cite {goity}: 
%%%%%%%%%%%%
\begin{align}
\left|S S_z;(p, q), Y, I I_z; S_c\right> = \sum\left(\begin{array}{cc} S_c & \frac{1}{2} \\ S_{cz} & s_z \end{array}\right|\left.\begin{array}{c} S \\ S_z \end{array}\right)\left(\begin{array}{cc} (p_c, q_c) & (1,0) \\ (Y_c,I_c,I_{cz}) & (y,i,i_z) \end{array}\right|\left.\begin{array}{c} (p, q) \\ (Y,I,I_z) \end{array}\right) \nn \\
\times\left|S_cS_{cz};(p_c, q_c),Y_c,I_cI_{cz}\right>\left|\frac{1}{2}s_z;(1,0),y,i,i_z\right>.
\end{align}
%%%%%%%%%%%%%%%%
Here $S$ is the total spin (for the four light quarks), $(p, q)$ are Dynkin indices describing the total irreducible representation of $SU(3)$, and $Y$ and $I$ are the total hypercharge and isospin.  The variables with the subscript $c$ refer to the corresponding quantities for the core, and the lower-case letters refer to the ``extra'' quark (which, of course, always has spin $\frac{1}{2}$ and $SU(3)$ representation $(1,0)$).  The second quantity in brackets is a Clebsch-Gordan coefficient for $SU(3)$, which can be written in terms of isoscalar factors and ordinary $SU(2)$ Clebsch-Gordan coefficients.  The sum runs over all possible values of $S_{cz}$, $s_z$, etc.  In the case of the ${\bf 15}_1$ states, we need to combine the two possible values of $S_c$ using the coefficients in Eq. (\ref{coef}).  

We can then use the Wigner-Eckart theorem to express each matrix element in terms of Clebsch-Gordan coefficients and the reduced matrix elements of $S_c^i$, $T_c^a$, $G_c^{ia}$, $s^i$, $t^a$, and $g^{ia}$.  The $SU(2)$ Clebsch-Gordan coefficients may be calculated in, e.g., Mathematica, or looked up in any of many published tables; analytic formulas for the necessary $SU(3)$ isoscalar factors appear in \cite{hecht,vergados}.\footnote{In case anyone actually wants to attempt this, it should be noted that there are a couple of typos in these otherwise very useful papers.  On page 31 of \cite{hecht}, the entry in Table 4 for $\epsilon_2\Lambda_2=-3\frac{1}{2}$, $\Lambda_1=\Lambda+\frac{1}{2}$, $\rho=2$ is incorrect: the factor $(\lambda+\mu+2-q+1)$ in the numerator should be $(\lambda+\mu+2-q)$, and the factor $(\mu+p-q)$ in the denominator should be $(\mu+p-q+1)$.  In Table 1 of \cite{vergados}, page 684, the entry for $\epsilon_2\Lambda_2=-1\frac{1}{2}$, $\Lambda_1=\Lambda+\frac{1}{2}$, $(\lambda^\prime, \mu^\prime)=(\lambda, \mu-1)$ is incorrect: the factor $(\mu-2)$ in the numerator should be $(\mu-q)$.}   One may also use the bosonic operator method described in \cite{manohar04}.  Explicit values for the relevant matrix elements appear in Appendix A.

The values of some of these matrix elements illustrate the tower structure.  In particular, the flavor ${\bf 3}$ states are in a different tower from the others.  The values of $\mathcal{O}_4$ for these two states differ at $\mathcal{O}(1)$ from those for the ${\bf \overline{6}}$, ${\bf 15}$ and ${\bf 15^\prime}$ states, so the two towers can have different masses as $N_c\to \infty$.  For this reason, it is not possible to construct a simple mass relation involving the ${\bf 3}_1-{\bf 3}_0$ splitting and the other heavy pentaquark multiplets.

There is, however, one  mass relation among the other negative-parity heavy pentaquarks:
%%%%%%%%%
\begin{equation} \label{mass1}
  {\bf 15}_2^P - {\bf 15}^{\prime P} _1 = 2({\bf \bar 6}_1^P - {\bf 15}_0^P) + \mathcal{O}(1/N_c^3).
\end{equation}
%%%%%%%%%
Note that this relation holds to order $1/N_c^2$. In addition to $O_5$ and $O_6$, the operator $O_{11} \equiv \frac{1}{N_c^3}S_c^2t^aT_c^a$ also contributes at $\mathcal{O}(1/N_c^2)$, and the relation remains true when this contribution is added.

The pentaquark states can also be related to the excited baryons, using the results from \cite{goity}, and to the ground state octet and decuplet baryons.  (Note that the core of the octet is a linear combination of $S_c =1$ and $S_c = 0$, with coefficients given in \cite{carlson99}.)  The mass relations include, 
%%%%%%%%%%%%%%%%%%%%%%%%%%%%%%%%%%%%%%%%%
\begin{align}
&{\bf \bar 6}_1^P - {\bf 15}_0^P =   \frac{2}{3}(\left < ^4{\bf 8}^E \right > - \left < ^2{\bf 10}^E \right >) + \mathcal{O}(1/N_c^2), \label{split1} \\ \nn \\
&({\bf 3}_1^P-{\bf 3}_0^P)-\frac{7}{11}({\bf 15}_2^P-{\bf 15}_1^P)+\frac{17}{11}({\bf 15}_1^P- {\bf 15}_0^P) = \nn \\
& \; \; \; \; \; \;  \; \; \; \; \frac{2}{11}( \left < ^2{\bf 10}^E \right >-\left < ^2{\bf 8}^E \right >) 
+\frac{4}{11}( \left < ^4{\bf 8}^E \right >-\left < ^2{\bf 8}^E \right >)+ \mathcal{O}(1/N_c^2),  \label{split2} \\ \nn \\
&\frac{11}{4}({\bf 15}^{\prime P}_1-{\bf 15}_2^P)+\frac{28}{11}({\bf 15}_2^P-{\bf 15}_1^P)-\frac{79}{11}({\bf 15}_1^P-{\bf \bar 6}_1^P)-2({\bf 15}_1^P-{\bf 3}_0^P)  = \nn \\ 
&  \; \; \; \; \; \;  \; \; \; \;({\bf 10}_{3/2}^N-{\bf 8}_{1/2}^N) 
+2\left < ^2{\bf 1}^E \right >+\frac{13}{11}\left < ^2{\bf 8}^E \right > -\frac{35}{11}\left < ^2{\bf 10}^E \right >  + \mathcal{O}(1/N_c^2). \label{split3}
\end{align}
%%%%%%%%%%%%%%%%%%%%%%%%%%%%
Here ${\bf 10}_{3/2}^N$ and ${\bf 8}_{1/2}^N$ are the ordinary octet and decuplet containing the ground-state nucleons and $\Delta$s.   $\left < ^2{\bf 10}^E \right >$, $\left < ^4{\bf 8}^E \right >$, $\left < ^2{\bf 8}^E \right >$, and $\left < ^2{\bf 1}^E \right >$ are spin averages of the excited baryons: 
%%%%%%%%%%%%%%%%%%%%%%%%%%%%%%
\begin{align}
  \left < ^2{\bf 10}^E \right > & =  \frac{1}{3}(^2{\bf 10}_{1/2}^E+2(^2{\bf 10}_{3/2}^E)) \nn \\
   \left < ^2{\bf 8}^E \right > & =  \frac{1}{3}(^2{\bf 8}_{1/2}^E+2(^2{\bf 8}_{3/2}^E)) \nn \\
    \left < ^2{\bf 1}^E \right > & =  \frac{1}{3}(^2{\bf 1}_{1/2}^E+2(^2{\bf 1}_{3/2}^E)) \nn \\
     \left < ^4{\bf 8}^E \right > & =  \frac{1}{6}(^4{\bf 8}_{1/2}^E+2(^4{\bf 8}_{3/2}^E)+3(^4{\bf 8}_{5/2}^E)).
\end{align}
%%%%%%%%%%%%%%%%%%%%%%%%%%%%
There is some ambiguity involved in identifying the ${\bf 8}_{1/2}^E$ and ${\bf 8}_{3/2}^E$ multiplets with physical states, because mixing presumably occurs, and the values of the mixing angles are not known with certainty.  The results here will be calculated assuming zero mixing.  We will identify the exited baryons as follows (choosing the non-strange states in the ${\bf 8}$ and ${\bf 10}$ cases):
%%%%%%%%%
\begin{align} \label{exbar}
^2{\bf 10}_{1/2}^E & = \Delta(1620) \nn \\
 ^2{\bf 10}_{3/2}^E & = \Delta(1600) \nn \\
   ^2{\bf 8}_{1/2}^E & =  N(1535) \nn \\
   ^2{\bf 8}_{3/2}^E & = N(1520) \nn \\
  ^4{\bf 8}_{1/2}^E & = N(1650) \nn \\
^4{\bf 8}_{3/2}^E & = N(1700) \nn \\
^4{\bf 8}_{5/2}^E & = N(1675) \nn \\
^2{\bf 1}_{1/2}^E & = \Lambda(1405) \nn \\
^2{\bf 1}_{3/2}^E & = \Lambda(1520). \nn \\
\end{align}
%%%%%%%%%

The order $N_c$ contribution to the pentaquark mass is about 1 GeV, so we estimate $\mathcal{O}(1/N_c^2)$ corrections to be of order 30 MeV, and $\mathcal{O}(1/N_c^3)$ to be of order 10 MeV.  Assuming no mixing and using the Particle Data Group values \cite{pdg} for the masses of the nonexotic baryons, we can give numerical estimates for the right-hand sides of Eqs. (\ref{split1}) - (\ref{split3}).  The mass difference $\left < ^4{\bf 8}^E \right > - \left < ^2{\bf 10}^E \right >$ is quite small, about 4 MeV; the error in Eq. (\ref{split1}) is estimated to be considerably larger than this.    Thus Eq. (\ref{split1}) indicates that the ${\bf \bar 6}_1^P$ and ${\bf 15}_0^P$ masses are close together, with a splitting of 4 $\pm$ 30 MeV, but cannot tell us which one is heavier.  The same applies to the ${\bf 15}_1^{\prime P}$ and ${\bf 15}_2^P$ masses, by Eq. (\ref{mass1}).  The right-hand side of Eq. (\ref{split2}) is about 83 MeV, again with an error of $\pm$ 30 MeV.  If all three pentaquark splittings on the left-hand side were equal, each would be about 40 $\pm$ 30 MeV.  Based on the estimate in \cite{stewart} that the isospin $\frac{1}{2}$ members of the ${\bf 3}_0^P$ should have mass 2580 MeV, this very rough guess suggests that the corresponding members of the ${\bf 3}_1^P$ would have mass 2620 $\pm$ 30 MeV, meaning that they would also be too light to decay to a $D_s$ plus a proton.    The right-hand side of Eq. (\ref{split3}) comes to -260 $\pm$ 30 MeV.

It is clear from comparing experimental evidence with expressions obtained in the large $N_c$ expansion that there is some mixing between the different ${\bf 8}$ states with the same spin and isospin.  Such mixing can be parameterized by two angles $\theta_{N_1}$ and $\theta_{N_3}$, constrained to be in the interval $[0, \pi)$ and defined by
%%%%%%%%
\begin{align}
\left( \begin{array}{c} N(1535) \\ N(1650) \end{array}\right) = \left(\begin{array}{cc} \cos{\theta_{N_1}} & \sin{\theta_{N_1}} \\ -\sin{\theta_{N_1}} & \cos{\theta_{N_1}} \end{array}\right)\left(\begin{array}{c} ^2{\bf 8}_{1/2}^E \\ ^4{\bf 8}_{1/2}^E \end{array}\right), \\ \nn \\
\left( \begin{array}{c} N(1520) \\ N(1700) \end{array}\right) = \left(\begin{array}{cc} \cos{\theta_{N_3}} & \sin{\theta_{N_3}} \\ -\sin{\theta_{N_3}} & \cos{\theta_{N_3}} \end{array}\right)\left(\begin{array}{c} ^2{\bf 8}_{3/2}^E \\ ^4{\bf 8}_{3/2}^E \end{array}\right).
\end{align}
%%%%%%%
Using data from the strong decays of the non-strange excited baryons, Goity, Schat, and Scoccola \cite{goity} estimate these angles to be
%%%%%%%%%
\begin{align} \label{angle1}
\theta_{N_1} & = 0.61 \pm 0.09, \nn \\
\theta_{N_3} & = 3.04 \pm 0.15.
\end{align}
%%%%%%%
Note that this value of $\theta_{N_3}$ is consistent with 0 mod $\pi$; $\theta_{N_1}$, however, may make a difference to our relations. 

On the other hand, in a later paper \cite{pirjol2} Pirjol and Schat find multiple different possibilities for the mixing angles, depending on the mass hierarchy of the three different towers to which the excited baryons are assigned.  A number of subsequent papers have attempted to narrow down the correct mixing angles based on empirical data.  Cohen and Lebed \cite{cohen2} find
%%%%%%%
\begin{align} \label{angle2}
\theta_{N_1} & \simeq 0.96, \nn \\
\theta_{N_3} & \simeq 2.72,
\end{align}
%%%%%%%
while a recent paper by Scoccola \cite{scoccola} holds that there is still some ambiguity, with
%%%%%%%%%
\begin{align} \label{angle3}
\theta_{N_1} & = 0.39 \pm 0.11, \nn \\
\theta_{N_3} & = (2.82 \; {\rm or} \; 2.38) \pm 0.11.
\end{align}
%%%%%%%
This wide range of results indicates that there is still a good deal of uncertainty on the subject, and we will not attempt to provide a definitive resolution here.  We will simply note that our numerical results may need some adjusting according to which mixing angles are chosen.  For example, the mass of the spin-averaged state $\left<^4{\bf 8}\right>$ appears in Eq. \eqref{split1}.  With no mixing, this mass is 1679 MeV, very close to the $\left<^2{\bf 10}\right>$ mass.  Using the mixing angles in \eqref{angle1}, we get 1672 MeV for the mass of the $\left<^4{\bf 8}\right>$, while with the values in \eqref{angle2}, we get 1656 MeV.  Using the two different cases in \eqref{angle3}, we get 1670 MeV and 1647 MeV respectively.  Thus the right-hand side of the relation \eqref{split1} could decrease by up to 32 MeV depending upon the choice of mixing angles.  This is not a large change, but it is somewhat important because it could cause the mass splitting to change sign.  However, as indicated above, the uncertainty in the relations is such that we cannot be sure about the sign in any case.  The relations \eqref{split1}-\eqref{split3} should be taken as rough guidelines rather than precise predictions.

\subsection{Heavy Quark Effective Theory}

Hadrons containing heavy quarks are simpler in some ways than their light counterparts.  Intuitively, a heavy quark, with mass $>>\Lambda_{QCD}$, is mostly untroubled by the ``brown muck'' inside a hadron; it is only lightly buffeted by the gluons and less-massive quarks zipping around it.  The simplifications due to the presence of heavy quarks have been rigorously quantified by a number of researchers; reference \cite{wiseheavy} explains heavy quark physics in detail.

Two types of symmetry characterize the interactions of heavy quarks in hadrons.  The first is heavy quark spin symmetry: in the limit $m_Q\to \infty$, a heavy quark's interactions with gluons are spin-independent, so arbitrary transformations can be made to the heavy quark's spin without changing the dynamics.  The second is heavy quark flavor symmetry: the heavy quark mass is clearly irrelevant as $m_Q\to\infty$, so one can also interchange heavy quark flavors without changing the dynamics.  The leading-order symmetry-breaking corrections for each of these symmetries are proportional to $1/m_Q$.

Hadrons containing a heavy quark have been studied using heavy quark effective theory (HQET).  The HQET Lagrangian,
%%%%%%%%%%%%%
\begin{align} \label{hqet}
\mathcal{L}_{HQET} = \bar{Q}_v(iv\cdot D)Q_v +\bar{Q}_v\frac{(iD)^2}{2m_Q}Q_v -Z_Q\bar{Q}_v\frac{gG_{\mu\nu}\sigma^{\mu\nu}}{4m_Q}Q_v+\mathcal{O}\left(\frac{1}{m_Q^2}\right), 
\end{align}
%%%%%%%%%%%%%%%%   
describes the interactions of a heavy quark $Q$ with velocity $v$ inside a hadron.  The velocity-dependent field $Q_v$ is related to the original quark field $Q$ by \cite{wiseheavy}
%%%%%%%%%%%
\begin{align}
Q(x) = e^{-im_Qv\cdot x}\left[Q_v(x) + \mathcal{Q}_v(x)\right], 
\end{align}
%%%%%%%%%%%%%%%
where $Z_Q$ is a renormalization factor with $Z_Q(\mu=m_Q)=1$, $Z_b/Z_c=\left[\alpha_s(m_b)/\alpha_s(m_c)\right]^{9/25}$, and 
%%%%%%%%%%%%
\begin{align}
Q_v(x) & = e^{im_Qv\cdot x}\frac{1+\xslash v}{2}Q(x), \nn \\
\mathcal{Q}_v(x) & = e^{im_Qv\cdot x}\frac{1-\xslash v}{2}Q(x).
\end{align}
%%%%%%%%%%%
This field redefinition absorbs the mass $m_Q$ of the heavy quark, so that there is no mass term in the Lagrangian \eqref{hqet}. 

The mass of a hadron with one heavy quark can be expanded in $1/m_Q$:
%%%%%%%%%%%%%%%%%%
\begin{align}
M(H_Q) = m_Q+\bar{\Lambda}-\frac{\lambda_1}{2m_Q} -d_H\frac{\lambda_2}{2m_Q}+\mathcal{O}\left(\frac{1}{m_Q^2}\right).
\end{align}
%%%%%%%%%%%%%%%%%%
Here $\bar{\Lambda}$ accounts for the mass of the light degrees of freedom, and $\lambda_1$ and $\lambda_2$ are defined by
%%%%%%%%%%%
\begin{align}
\lambda_1 & = \left<H_Q(v)\right|\bar{Q}_v(iD)^2Q_v\left|H_Q(v)\right>, \\
\label{l2} d_H\lambda_2 & = \frac{1}{2}Z_Q \left<H_Q(v)\right|\bar{Q}_vgG_{\mu\nu}\sigma^{\mu\nu}Q_v\left|H_Q(v)\right>.
\end{align}
%%%%%%%%%%%%%
In Eq. \eqref{l2}, the Clebsch factor $d_H$ is equal to $-4(J_\ell \cdot J_Q)$.
The parameters $\bar{\Lambda}$, $\lambda_1$, and $\lambda_2$ differ depending on whether the hadron in question is a baryon or a meson; here we will be focusing on baryons.

Heavy quark effective theory can be combined with the large $N_c$ formalism to produce an expansion in $1/N_c$ and $1/m_Q$ \cite{jenkins}.  In the limit $m_b\to \infty$, $m_c\to\infty$, $N_c\to \infty$, holding $m_c/m_b$ and $N_c\Lambda_{QCD}/m_b$ fixed, there is a combined light-quark and heavy-quark spin-flavor symmetry $SU(6)_\ell \times SU(4)_h$.  The breaking of this symmetry can be parameterized as in the previous section, except that now we must include the heavy quark spin operator $J_Q^i$, which contributes at order $1/m_Q$.

In the heavy pentaquarks we consider here, the light quarks combine with the heavy antiquark to produce the twelve states   $^3{\bf 15}^{\prime P} _{1/2}$, $^3{\bf 15}^{\prime P} _{3/2}$, $^5{\bf 15} _{3/2}^P$, $^5{\bf 15} _{5/2}^P$,  $^3{\bf 15} _{1/2}^P$, $^3{\bf 15} _{3/2}^P$,  $^1{\bf 15} _{1/2}^P$,  $^3{\bf \bar 6} _{1/2}^P$, $^3{\bf \bar 6} _{3/2}^P$, $^3{\bf 3} _{1/2}^P$, $^3{\bf 3} _{3/2}^P$, and  $^1{\bf 3} _{1/2}^P$.  (The notation here is $^{2j_{\ell}+1}{\bf F}_J$, where $j_{\ell}$ is the light-quark spin, ${\bf F}$ is the flavor representation, and $J$ is the total spin of the state.)    The singlet operators at order $1/(N_c m_Q)$ are
%%%%%%%%%%%%%%%%%
\begin{align}
O_7 & \equiv \frac{1}{N_cm_Q} S_c^i J_Q^i \nn \\
O_8 & \equiv \frac{1}{N_cm_Q} s^i J_Q^i \nn \\
O_9 & \equiv \frac{2}{N_c^2m_Q} t^a \{J_Q^i,G_c^{ia} \} \nn \\
O_{10} & \equiv \frac{1}{N_c^2m_Q} g^{ia} J_Q^i T_c^a,
\end{align}
%%%%%%%%%%%%%%%%%%%%%%%%%%%%%
where $J_Q^i$ is the spin of the heavy antiquark. Matrix elements of these operators appear in Table A.2.  We find the mass relations
%%%%%%%%%%%%%%
\begin{align}
^3{\bf \bar 6} _{3/2}^P - ^3{\bf \bar 6} _{1/2}^P & =  \frac{9}{10}(^5{\bf 15} _{5/2}^P - ^5{\bf 15} _{3/2}^P) - \frac{1}{2}(^3{\bf 15}^{\prime P} _{3/2} - ^3{\bf 15}^{\prime P} _{1/2}) + \mathcal{O}(1/N_c^2m_Q),  \\
^5{\bf 15} _{5/2}^P - ^5{\bf 15} _{3/2}^P & = \frac{5}{3}(^3{\bf 15}^{P} _{3/2} - ^3{\bf 15}^{P} _{1/2})  + \mathcal{O}(1/N_c^2m_Q), \\
^3{\bf \bar 6} _{3/2}^P - ^3{\bf \bar 6} _{1/2}^P & =  {\bf 6} _{3/2}^N - {\bf 6} _{1/2}^N
 + \mathcal{O}(1/N_c^2m_Q), \label{heavy}
\end{align}
%%%%%%%%%%%%%%%%%%%%%%%%%%%%%%%%%%%%%%%%
where ${\bf 6} _{1/2}^N$ and ${\bf 6} _{3/2}^N$ are the non-exotic heavy baryon  multiplets containing the $\Sigma_{c,b}$ and $\Sigma_{c,b}^*$, respectively.  In the charmed case, the masses of the  $\Sigma_{c}$ and $\Sigma_{c}^*$ are 2455 MeV and 2520 MeV respectively, so the mass splitting in Eq. (\ref{heavy}) is 65 MeV \cite{pdg}.  For the bottom case, mass measurements of the $\Sigma_{b}$ and $\Sigma_{b}^*$ are not currently available.

States with the same spin and flavor quantum numbers may mix.  There are three mixing angles, for the two ${\bf 3}_{1/2}$ states, the two ${\bf 15}_{1/2}$ states, and the two ${\bf 15}_{3/2}$ states.  Each pair mixes at order $1/(N_cm_Q)$; off-diagonal matrix elements corresponding to the mixing can also be found in Table A.2.

\subsection{$SU(3)$ Breaking}

The breaking of $SU(3)$ flavor symmetry can be quantified by the parameter $\epsilon \simeq 0.3$.  $SU(3)$ is broken at order $\epsilon$ by octet operators, at order, $\epsilon^2$ by operators in the ${\bf 27}$ representation, and at order $\epsilon^3$ by operators in the ${\bf 64}$.  By calculating the matrix elements of these operators, one can find relations between different flavor states, such as the famous Gell-Mann-Okubo formula for the mass splittings among the octet baryons.

$SU(3)$ breaking can be combined with the $1/N_c$ expansion to produce a hierarchy for baryon mass splittings \cite{jenkinssu3}. For example, the baryon mass relations 
%%%%%%%%%%
\begin{align} \label{su31}
\frac{5}{2}(6N-3\Sigma+\Lambda-4\Xi)-(2\Delta-\Xi^*-\Omega)=0,
\end{align}
%%%%%%%%%%%%%
 and 
%%%%%%%%%%%%%
\begin{align} \label{su32}
\frac{1}{3}(N-3\Sigma+\Lambda+\Xi)=0,
\end{align}
%%%%%%%%%%%%%%%%
 are both order $\epsilon$ in $SU(3)$ breaking, but the former is order $N_c^0$ while the latter is order $1/N_c$ in the large $N_c$ expansion.  Experimentally, the relation (\ref{su31}) is broken at the 20\% level, while (\ref{su32}) is broken only at the 6\% level.  The relation
%%%%%%%%%%%%%
\begin{align}
\frac{1}{2}(-2N-9\Sigma+3\Lambda+8\Xi)+(2\Delta -\Xi^*-\Omega)=0,
\end{align}
%%%%%%%%%%%%%%% 
is order $\epsilon/N_c^2$; it is good to the 1\% level.  This is a good example of the usefulness of the $1/N_c$ expansion for describing and predicting the baryon mass spectrum; before $1/N_c$ calculations were done, it was not understood why some of these relations worked so much better than others.

In the case of negative-parity pentaquarks, there are three octet operators that break $SU(3)_f$ at $\mathcal{O}(\epsilon N_c^0)$ \cite{goity, pirjolschat}\footnote{Ref. \cite{goity} includes a fourth operator that depends on the orbital angular momentum $\ell$; this is irrelevant here because $\ell=0$ for the states we are considering.}:
%%%%%%%%%
\begin{align}
B_1 & = t^8 \nn \\
B_2 & = T_c^8 \nn \\
B_3 & = \frac{1}{N_c}d^{8ab}g^{ia}G_c^{ib}.
\end{align}
%%%%%%%
Here $d^{abc}$ is the completely symmetric $SU(3)$ coefficient; its relevant nonzero values are
%%%%%%%%
\begin{align}
d^{811} =d^{822} = d^{833} &=  \frac{1}{\sqrt{3}} \nn \\
d^{844}  = d^{855} = d^{866} = d^{877} &= -\frac{1}{2\sqrt{3}} \nn \\
d^{888}  &= -\frac{1}{\sqrt{3}}.
\end{align}
%%%%%%%%
Explicit values for the matrix elements of these three operators are given in Appendix A.   The operator $B_2$ is actually of order $\epsilon N_c$ here; however, the order $N_c$ part has the same value, $\frac{N_c}{2\sqrt{3}}$, for each of the seven multiplets, so it does not affect the splittings.

By considering empirical evidence, reference \cite{goity} finds that in the case of the excited baryons in the ${\bf 70}$ of $SU(6)$, the operator $B_3$ is weak and does not contribute much to the $SU(3)$ breaking.  This may or may not be true in the case of heavy pentaquarks.  The contributions from $B_3$ are listed in Table A.4.  

From Tables A.3 and A.4, we see that within the ${\bf \overline{6}}_1$ multiplet, the masses of states with different strangeness are split only at order $\epsilon/N_c \sim 10\%$, and the three different masses are equally spaced.  If we label the states by their $S$ values as ${\bf \overline{6}}_1(0)$, ${\bf \overline{6}}_1(-1)$, and ${\bf \overline{6}}_1(-2)$, then we have the relation
%%%%%%%%%%%%%%%%%%
\begin{align} \label{equal}
M({\bf \overline{6}}_1(-1))-M({\bf \overline{6}}_1(0)) = M({\bf \overline{6}}_1(-2))-M({\bf \overline{6}}_1(-1)) +\mathcal{O}(\epsilon/N_c^2).
\end{align}
%%%%%%%%%%%%%%%%

For the other six multiplets, the corresponding mass splittings are order $\epsilon \sim 30\%$, because of the operator $B_2$.  Equal-spacing relations similar to Eq. \eqref{equal} can be derived for the other multiplets as well, though they will be satisfied only to order $\epsilon$.  (Of course, in the case of the ${\bf 3}$ multiplets, such a relation is not terribly useful, because each has only two possible combinations of strangeness and isospin, and thus one mass splitting.)  In the ${\bf 15}$ multiplets, there are some states with the same strangeness but different isospin: $S=-1$ states can have $I=3/2$ or $I=1/2$, and $S=-2$ states states can have $I=1$ or $I=0$.  (This is illustrated in the diagram in Figure \ref{fig:su3}; the circles surrounding some states indicate degeneracies.)  The mass splittings between states with the same $S$ but different $I$ are all of order $\epsilon/N_c$.  

Pirjol and Schat \cite{pirjolschat} predict that the two isospin multiplets with $(I, J_q)=(1/2,2)$ and $(3/2,0)$, some of  the $S=-1$ members of the ${\bf 15}_0$ and ${\bf 15}_2$ multiplets, will be split only at $\mathcal{O}(1/N_c)$ to all orders in $SU(3)$ breaking.  This is a result of the fact that those two multiplets are in the same tower, labeled by $K=3/2$.  The explicit matrix elements listed in Table A.3 confirm the prediction for the order $\epsilon$ $SU(3)$ breaking.  This can be seen by expanding each matrix element to order $N_c^0$:
%%%%%%%%
\begin{align} \label{same}
\left<{\bf 15}_0\right|(c_1B_1+c_2B_2+c_3B_3)\left|{\bf 15}_0 \right> & =  \frac{1}{2\sqrt{3}}c_1+\left(\frac{N_c}{2\sqrt{3}}+\frac{3S}{2\sqrt{3}}\right)c_2 -\frac{1}{16\sqrt{3}}c_3+\mathcal{O}(1/N_c), \nn \\
\left<{\bf 15}_2\right|(c_1B_1+c_2B_2+c_3B_3)\left|{\bf 15}_2 \right> & =  \frac{1}{2\sqrt{3}}c_1+\left(\frac{N_c}{2\sqrt{3}}+\frac{3S}{2\sqrt{3}}\right)c_2-\frac{1}{16\sqrt{3}}c_3+\mathcal{O}(1/N_c).
\end{align}
 %%%%%%%
At order $1/N_c$, however, the matrix elements depend on both strangeness and isospin in the combination $A=4I^2+4I-S^2$, so the $(I, J_q)=(1/2,2)$ and $(3/2,0)$ states will be split at this order.

It so happens that the $\mathcal{O}(\epsilon)$ values of $B_1$, $B_2$, $B_3$ given in Eq. \eqref{same} are the same for the ${\bf 15}^\prime_1$ and ${\bf 15}_1$ as well.    Therefore, the matching-strangeness states from any of these multiplets are split only at order $\epsilon/N_c$.  At $S=-1$, this includes the states with $(I, J_q)$ values $(1/2,0)$, $(3/2,1)$, and $(3/2,2)$ as well as those listed above.  Presumably some of this degeneracy will be broken at higher orders in $SU(3)$ breaking.

We can also derive some relations between the mass splittings in different multiplets. For example (again denoting states by their strangeness $S$), 
%%%%%%%%%
\begin{align} \label{pairs}
M({\bf 3}_0(-2))-M({\bf 3}_0(-1)) & = M({\bf 3}_1(-2))-M({\bf 3}_1(-1)) \nn \\
&= M({\bf 15}_0(-2))-M({\bf 15}_0(-1)) \nn \\
& = M({\bf 15}_2(-2))-M({\bf 15}_2(-1)) \nn \\
& = M({\bf 15}^\prime_1(-2))-M({\bf 15}^\prime_1(-1)) + \mathcal{O}(\epsilon/N_c).
\end{align}
%%%%%%%%%
(Eq. \eqref{pairs} holds for either isospin value of the ${\bf 15}_0(S)$ and ${\bf 15}_2(S)$ states.)

So far, we have ignored mixing among the various multiplets.  States in the ${\bf 15}_2$ will not mix, because there are no other multiplets with spin 2, but for the other states, mixing can potentially be an important effect.  The following table shows the values of isospin and strangeness for which mixing can potentially occur between the different multiplets:
%%%%%%%%%
\begin{align} \label{mixing}
\begin{array}{cc}  {\rm Multiplets} & (I, S) \\
{\bf 3}_0, {\bf 15}_0 & (1/2, -1);  \; (0, -2) \\
{\bf 3}_1, {\bf \overline{6}}_1 & (1/2, -1);  \\
{\bf 3}_1, {\bf 15}_1 & (1/2, -1);  \; (0, -2) \\
{\bf \overline{6}}_1, {\bf 15}_1 & (1/2, -1);  \; (1, -2) \\
{\bf 15}^\prime_1, {\bf 15}_1 & (3/2, -1);  \; (1, -2); \; (1/2, -3).  
\end{array}
\end{align}
%%%%%%%%%%

As noted in \cite{pirjolschat}, the tower structure provides some information about  the pattern of $SU(3)$ breaking.  For example, there will be mixing among the states with $S=-1$ and $(I, J_q)=(1/2, 0)$, from the ${\bf 3}_0$ and the ${\bf 15}_0$, even though these states are in different towers.   However, in the large $N_c$ limit, the eigenvalues of the mass matrix must coincide with the masses of the corresponding tower states.  The same can be said for mixing between the ${\bf 3}_1$ and the ${\bf \overline{6}}_1$, ${\bf 15}_1$, and ${\bf 15}^\prime_1$.

The matrix elements of $B_1$ and $B_2$ corresponding to mixing are listed in the last five rows of Table A.3.  It is interesting to note that in each case, $\left<B_2\right>=-\left<B_1\right>$.  However, this does not necessarily mean that the two contributions cancel; $B_1$ and $B_2$ in general have different coefficients in the effective Hamiltonian, and these coefficients may not even have the same sign.  The mixing from the three coefficients is of order $\epsilon/\sqrt{N_c}$ for the first three cases listed in \eqref{mixing}, and of order $\epsilon/N_c$ for the last two.  

\newpage

\subsection{Heavy Quark, Light Antiquark States}

\begin{figure}[h]
\PSbox{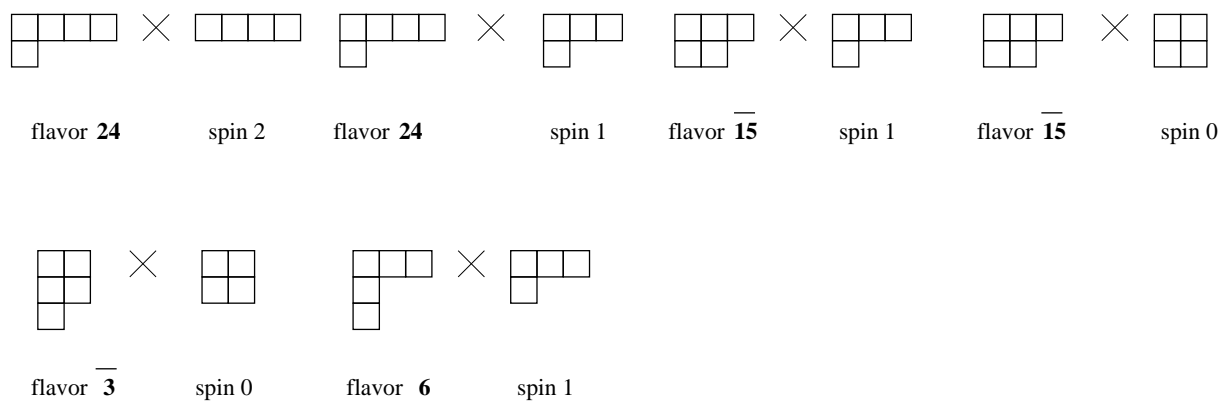 hoffset=10 voffset=10 hscale=55
vscale=55}{6.8in}{3.0in}
\caption[Spin-flavor decomposition for positive-parity $\bar{q}Qqqq$ states]{Young tableaux showing the flavor $SU(3)$ $\times$ spin $SU(2)$ decomposition of the positive-parity $\bar{q}Qqqq$ pentaquarks.  The two representations in the lower row contain no exotic states.}
\label{fig:spinflavorQ+}
\end{figure}

The states of the form $\bar{q}Qqqq$ discussed in Chapter 2 can also be considered in the context of large $N_c$.  Positive-parity, $P$-wave states of this type come from a completely symmetric representation of $SU(6)$.  This representation decomposes into the spin-flavor representations ${\bf 24}_2$,  ${\bf 24}_2$, ${\bf \overline{15}}_1$, and ${\bf \overline{15}}_0$, which contain exotic states, plus ${\bf \overline{3}}_0$ and ${\bf 6}_1$, which are not exotic.  These are illustrated in Figure \ref{fig:spinflavorQ+}.

%%%%%%%%%%%%%%%%%%%%%%%%%%%%%%
%%%%%%%%%%%%%%%%%%%%%%%%%%%%%%%

\begin{figure}[t]
\PSbox{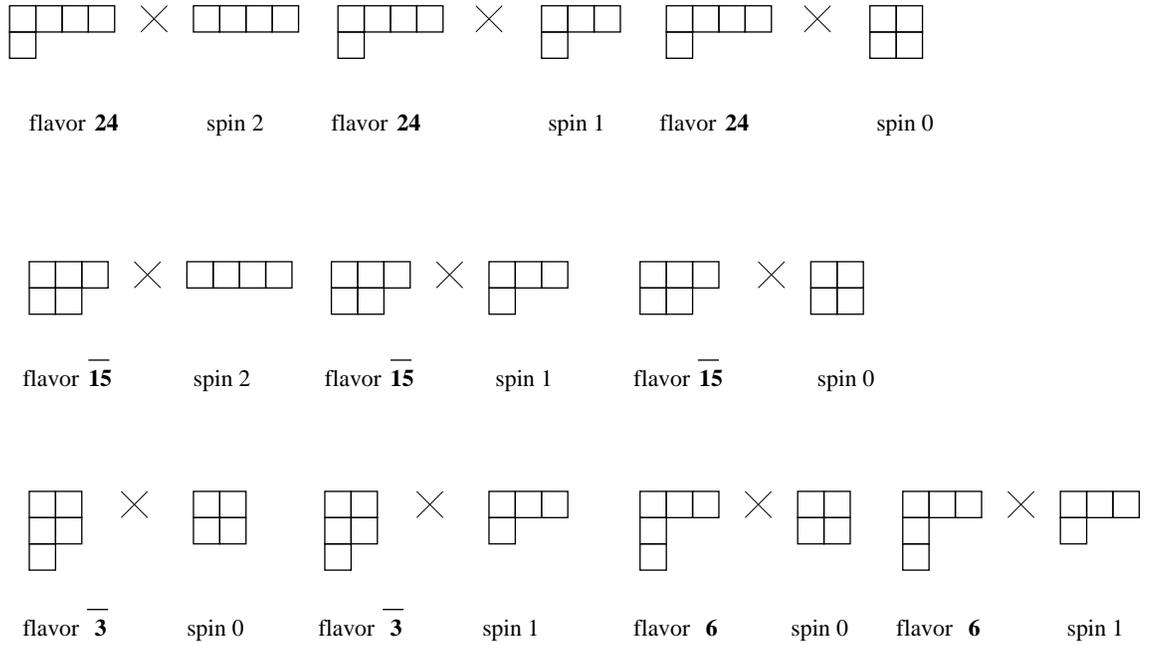 hoffset=10 voffset=5 hscale=55
vscale=55}{6.8in}{4.0in}
\caption[Spin-flavor decomposition for negative-parity $\bar{q}Qqqq$ states]{Young tableaux showing the flavor $SU(3)$ $\times$ spin $SU(2)$ decomposition of the negative-parity $\bar{q}Qqqq$ pentaquarks.  The four representations in the last row contain no exotic states.}
\label{fig:spinflavorQ-}
\end{figure}

The negative-parity, $S$-wave pentaquarks with a heavy quark come from a mixed-symmetry representation of $SU(6)$, whose spin-flavor decomposition is shown in Figure \ref{fig:spinflavorQ-}.  Here we get quite a few more multiplets: ${\bf 24}_2$, ${\bf 24}_1$, ${\bf 24}_0$,  ${\bf \overline{15}}_2$, ${\bf \overline{15}}_1$, and ${\bf \overline{15}}_0$, which contain some exotic states, plus the nonexotic ${\bf \overline{3}}_1$, ${\bf \overline{3}}_0$, ${\bf 6}_1$, and ${\bf 6}_0$.  

%%%%%%%%%%%%%%%%%%%%%%%%%%%%%%
%%%%%%%%%%%%%%%%%%%%%%%%%%%%%%%
The ${\bf \overline{15}}_0$ is the multiplet considered in Chapter 2. It can be constructed from ``good'' antisymmetric diquarks (if we allow one diquark to contain a heavy quark).  The other negative-parity states can be built from combinations of ``good'' and ``bad'' diquarks.  

The large $N_c$ formalism could be applied to these negative-parity states as well, but doing so would probably not be worth the trouble.  In this case, we would have to treat three particles as ``special:'' the heavy quark, the antiquark, and the ``extra'' quark not included in the core.  So many different operators would need to be included that it would be unlikely for any useful relations to emerge.

\section{Summary and Discussion}

It should be emphasized that the large $N_c$ expansion makes no prediction as to the existence of pentaquarks or other exotic states.  Nothing in this thesis can settle the experimental question of whether the $\Theta^+$ is real, or whether it has any heavy counterparts.  What we have done is to outline a method of studying these exotic baryons, should they prove to be of physical interest. 

In this chapter, we have generalized the negative-parity pentaquarks introduced in Chapter 2.  By moving from a specific constituent quark model, the diquark model, to the broader context of large $N_c$, we have uncovered a whole host of other possible negative-parity heavy pentaquarks.  As with the $F$ and $F_b$ states described in Section 2.2.3, it is far from certain whether any of these other pentaquarks are stable against strong decay, or whether they have other properties causing them to be narrow enough to be observable.  Nevertheless, we have at least shown how they can be accommodated in the large $N_c$ picture, and derived some relations that they must obey if they do indeed exist.  Of particular note is the prediction that the ${\bf 3}_1$ states, like the ${\bf 3}_0$ states considered in Chapter 2, may also be light enough to be stable against strong decay.  

At this point, it is up to the experimentalists to determine whether the $\Theta^+$ is real or an unfortunate artifact of misguided data analysis.  As noted in the introduction, planned experiments at CLAS and COSY-TOF will hopefully settle this question soon.  If the $\Theta^+$ does turn out to exist, this thesis points to some interesting places to look for other pentaquark states.

\appendix

\chapter{Explicit Matrix Elements}

\section{Matrix Elements of Light Operators}

\begin{table}[h]
\renewcommand{\arraystretch}{2.0}
\begin{tabular}{|| c | c | c | c | c | c | c | c | c |}
  \hline
  &  $O_1$ &  $O_2$ & $O_3$ &  $O_4$ &  $O_5$ &  $O_6$ & $O_{11}$ \\ 
 & $N_c {\bf 1}$ & $\frac{1}{N_c}S_c^2$ &  $\frac{1}{N_c}s^iS_c^i$ & $\frac{1}{N_c}t^aT_c^a$ &  $\frac{1}{N_c^2}t^a \{ S_c^i, G_c^{ia} \}$ &  $\frac{1}{N_c^2}g^{ia}S_c^iT_c^a$ & $\frac{1}{N_c^3}S_c^2t^aT_c^a$ \\ \hline 
 ${\bf 3}_0^P$ & $N_c$ & $ \frac{3}{4N_c}$ & $- \frac{3}{4N_c}$ & $- \frac{N_c+6}{6N_c}$ & $- \frac{1}{2N_c^2}$ & $ \frac{N_c+6}{8N_c^2}$ & $-\frac{1}{8N_c^2}$ \\ \hline
  ${\bf 3}_1^P$ & $N_c$ & $ \frac{3}{4N_c}$ & $\frac{1}{4N_c}$ & $- \frac{N_c+6}{6N_c}$ & $- \frac{1}{2N_c^2}$ & $- \frac{N_c+6}{24N_c^2}$ & $-\frac{1}{8N_c^2}$ \\ \hline
 ${\bf \bar 6}^P_1$ & $N_c$ & $ \frac{3}{4N_c}$ & $ \frac{1}{4N_c}$ & $\frac{N_c-9}{12N_c}$ & $- \frac{3N_c+5}{4N_c^2}$ & $ \frac{N_c-9}{48N_c^2}$ &  $\frac{1}{16N_c^2}$ \\ \hline
${\bf 15}_0^P$ &  $N_c$ & $ \frac{3}{4N_c}$ & $- \frac{3}{4N_c}$ & $\frac{N_c+3}{12N_c}$ & $\frac{N_c+3}{4N_c^2}$ & $- \frac{N_c+3}{16N_c^2}$ & $\frac{1}{16N_c^2}$ \\ \hline
${\bf 15}_1^P$ & $N_c$ & $ \frac{7N_c+16}{4N_c^2}$ & $- \frac{7N_c+16}{12N_c^2}$ & $\frac{N_c^2-3N_c-25}{12N_c^2}$ & $- \frac{3N_c+23}{24N_c^2}$ & $\frac{-N_c+19}{48N_c^2}$ &  $\frac{7}{48N_c^2}$ \\ \hline
${\bf 15}_2^P$ &  $N_c$ & $ \frac{15}{4N_c}$ & $ \frac{3}{4N_c}$ & $\frac{N_c-15}{12N_c}$ & $- \frac{5(N_c+1)}{4N_c^2}$ & $ \frac{N_c-15}{16N_c^2}$ &  $\frac{5}{16N_c^2}$ \\ \hline
${\bf 15}^{\prime P} _1$ & $N_c$ & $ \frac{15}{4N_c}$ & $- \frac{5}{4N_c}$ & $\frac{N_c+9}{12N_c}$ & $\frac{3N_c+11}{4N_c^2}$ & $- \frac{5(N_c+9)}{48N_c^2}$ &  $\frac{5}{16N_c^2}$ \\ \hline
${\bf 10}_{3/2}^N$ & $N_c$ & $ \frac{2}{N_c}$ & $ \frac{1}{2N_c}$ & $\frac{N_c+5}{12N_c}$ & $\frac{3N_c+7}{6N_c^2}$ & $- \frac{N_c+5}{24N_c^2}$ &  $\frac{1}{6N_c^2}$ \\ \hline
${\bf 8}_{1/2}^N$ & $N_c$ & $ \frac{3(N_c-1)}{2N_c^2}$ & $- \frac{3(N_c-1)}{4N_c^2}$ & $\frac{N_c^2-10N_c+9}{12N_c^2}$ & $\frac{-3N_c^2+2N_c}{8N_c^3}$ & $ \frac{N_c^2+14N_c}{16N_c^3}$ & $\frac{1}{8N_c^2}$ \\ \hline
\end{tabular}
\caption[Matrix elements of light operators]{Matrix elements of singlet operators $O_1$ through $O_6$ and $O_{11}$ to order $1/N_c^2$. It should be noted that the {\it exact} matrix elements for the ${\bf 15}_1^P$ multiplet are in agreement with the results of \cite{pirjolschat}.  For Table I, $\frac{1}{N_c(N_c+1)}$ was expanded as $\frac{1}{N_c^2}-\frac{1}{N_c^3}+...$, and only terms up to order $1/N_c^2$ were kept.}
\end{table}

\newpage

\section{Matrix Elements of Heavy Operators}

\begin{table}[h]
\renewcommand{\arraystretch}{1.7}
\begin{tabular}{|| c | c | c | c | c |}
  \hline
 & $O_7$ & $O_8$ & $O_9$ & $O_{10}$ \\ 
& $\frac{1}{N_cm_Q}S_c^iJ_Q^i$ &  $\frac{1}{N_cm_Q}s^iJ_Q^i$ & $\frac{2}{N_c^2m_Q}t^a \{ J_Q^i, G_c^{ia} \}$ & $\frac{1}{N_c^2m_Q}g^{ia}J_Q^iT_c^a$ \\ \hline
$^1{\bf 3}_{1/2}^P$ & 0 & 0 & 0 & 0  \\ \hline
$^3{\bf 3}_{3/2}^P$ & $\frac{1}{4N_cm_Q}$ & $\frac{1}{4N_cm_Q}$ & 0  & $- \frac{1}{24N_cm_Q}$ \\ \hline
$^3{\bf 3}_{1/2}^P$ & $- \frac{1}{2N_cm_Q}$ & $- \frac{1}{2N_cm_Q}$ & 0  & $\frac{1}{12N_cm_Q}$ \\ \hline
$^3{\bf \bar 6}_{3/2}^P$ & $\frac{1}{4N_cm_Q}$ & $\frac{1}{4N_cm_Q}$ & $- \frac{1}{4N_cm_Q}$  & $\frac{1}{48N_cm_Q}$ \\ \hline
$^3{\bf \bar 6}_{1/2}^P$ & $- \frac{1}{2N_cm_Q}$ & $- \frac{1}{2N_cm_Q}$ &  $\frac{1}{2N_cm_Q}$  & $- \frac{1}{24N_cm_Q}$ \\ \hline
$^1{\bf 15}_{1/2}^P$ & 0 & 0 & 0 & 0  \\ \hline
$^3{\bf 15}_{3/2}^P$ & $\frac{3}{8N_cm_Q}$ & $\frac{1}{8N_cm_Q}$ & $- \frac{1}{8N_cm_Q}$  & $\frac{1}{96N_cm_Q}$ \\ \hline
$^3{\bf 15}_{1/2}^P$ & $- \frac{3}{4N_cm_Q}$ & $-\frac{1}{4N_cm_Q}$ & $\frac{1}{4N_cm_Q}$  & $- \frac{1}{48N_cm_Q}$ \\ \hline
$^5{\bf 15}_{5/2}^P$ & $\frac{3}{4N_cm_Q}$ & $\frac{1}{4N_cm_Q}$ & $- \frac{1}{4N_cm_Q}$  & $\frac{1}{48N_cm_Q}$ \\ \hline
$^5{\bf 15}_{3/2}^P$ & $- \frac{9}{8N_cm_Q}$ & $- \frac{3}{8N_cm_Q}$ & $\frac{3}{8N_cm_Q}$  & $- \frac{1}{32N_cm_Q}$ \\ \hline
$^3{\bf 15}^{\prime P} _{3/2}$ & $\frac{5}{8N_cm_Q}$ & $- \frac{1}{8N_cm_Q}$ & $\frac{1}{8N_cm_Q}$  & $- \frac{1}{96N_cm_Q}$ \\ \hline
$^3{\bf 15}^{\prime P} _{1/2}$ & $- \frac{5}{4N_cm_Q}$ & $\frac{1}{4N_cm_Q}$ & $- \frac{1}{4N_cm_Q}$  & $\frac{1}{48N_cm_Q}$ \\ \hline
${\bf 6}_{3/2}^N$ & $\frac{1}{4N_cm_Q}$ & $\frac{1}{4N_cm_Q}$ & $- \frac{1}{4N_cm_Q}$  & $ \frac{1}{48N_cm_Q}$ \\ \hline
${\bf 6}_{1/2}^N$ & $- \frac{1}{2N_cm_Q}$ & $- \frac{1}{2N_cm_Q}$ &  $\frac{1}{2N_cm_Q}$  & $- \frac{1}{24N_cm_Q}$ \\ \hline
$^3{\bf 3}_{1/2}^P-^1{\bf 3}_{1/2}^P$ & $-\frac{\sqrt{3}}{4N_cm_Q}$ & $\frac{\sqrt{3}}{4N_cm_Q}$ & 0 & $-\frac{1}{2\sqrt{3}N_cm_Q}$ \\ \hline
$^3{\bf 15}_{1/2}^P-^1{\bf 15}_{1/2}^P$ & $\frac{1}{2\sqrt{2}N_cm_Q}$ &  $-\frac{1}{2\sqrt{2}N_cm_Q}$ & $-\frac{1}{2\sqrt{2}N_cm_Q}$  & $-\frac{1}{24\sqrt{2}N_cm_Q}$ \\ \hline
$^5{\bf 15}_{3/2}^P-^3{\bf 15}_{3/2}^P$ & $-\frac{\sqrt{5}}{8N_cm_Q}$ & $\frac{5}{8\sqrt{3}N_cm_Q}$ & $\frac{\sqrt{5}}{6N_cm_Q}$  & $\frac{\sqrt{5}}{96N_cm_Q}$  \\ \hline
\end{tabular}
\caption[Matrix elements of heavy operators]{Matrix elements of heavy operators $O_7$ through $O_{10}$ to order $1/N_cm_Q$.  The last three rows show off-diagonal matrix elements, which are related to the mixing angles.  All results in this table are expanded to $\mathcal{O}(1/(N_cm_Q)$.}
\end{table}

\newpage

%\begin{landscape}
\section{Matrix Elements of $SU(3)$ Breaking Operators $B_1$, $B_2$}

\begin{table}[h]
\renewcommand{\arraystretch}{2.2}
\begin{tabular}{|| c | c | c | }
  \hline
  &  $B_1$ &  $B_2$ 
  \\ 
 & $t^8$ & $T^8_c$  \\ \hline 
 ${\bf 3}_0^P$ & $-\frac{\left(N_c^2+3N_c(S+1)-(3S+10)\right)}{\sqrt{3}(N_c+1)(N_c+5)}$ & $\frac{N_c^3+3N_c^2(S+3)+N_c(24S+17)}{2\sqrt{3}(N_c+1)(N_c+5)}$     \\ \hline
  ${\bf 3}_1^P$ & $-\frac{\left(N_c^2+3N_c(S+1)-(3S+10)\right)}{\sqrt{3}(N_c+1)(N_c+5)}$ & $\frac{N_c^3+3N_c^2(S+3)+N_c(24S+17)}{2\sqrt{3}(N_c+1)(N_c+5)}$ 
     \\ \hline
 ${\bf \bar 6}^P_1$ & $\frac{N_c+6S-5}{2\sqrt{3}(N_c+1)}$ & $\frac{N_c^2+N_c-6S+6}{2\sqrt{3}(N_c+1)}$   \\ \hline
${\bf 15}_0^P$ & $\frac{4N_c+3A+18S-4}{8\sqrt{3}(N_c+5)}$  & $\frac{4N_c^2+4N_c(3S+5)+3(14S+8)-3A}{8\sqrt{3}(N_c+5)}$    \\ \hline
${\bf 15}_1^P$ & $\frac{8N_c^3+N_c^2(3A+30S-24)}{16\sqrt{3}N_c^3}$  & $\frac{24N_c^4+N_c^3(72S)-N_c^2(9A+90S+527)}{48\sqrt{3}N_c^3}$  \\ \hline
${\bf 15}_2^P$ &  $\frac{8N_c-3A-18S+16}{16\sqrt{3}(N_c+1)}$ & $\frac{8N_c^2 +8N_c(3S+1)+3A-3(14S+8)}{16\sqrt{3}(N_c+1)}$   \\ \hline
${\bf 15}^{\prime P} _1$ & $\frac{2N_c+15S+14}{4\sqrt{3}(N_c+7)}$ & $\frac{2N_c^2+2N_c(3S+7)+27S}{4\sqrt{3}(N_c+7)}$    \\ \hline
 ${\bf 3}_0^P-{\bf 15}_0^P$ &  $-\frac{3}{2}\frac{1}{N_c+5}\sqrt{\frac{(-S)(N_c-1)(N_c+2S+5)}{N_c+1}}$  & $\frac{3}{2}\frac{1}{N_c+5}\sqrt{\frac{(-S)(N_c-1)(N_c+2S+5)}{N_c+1}}$  \\ \hline
 ${\bf 3}_1^P-{\bf \overline{6}}_1^P$ & $\frac{\sqrt{3}}{2}\frac{1}{N_c+1}\sqrt{\frac{(N_c+3)(N_c-1)}{N_c+5}}$ &  $-\frac{\sqrt{3}}{2}\frac{1}{N_c+1}\sqrt{\frac{(N_c+3)(N_c-1)}{N_c+5}}$\\ \hline
 ${\bf 3}_1^P-{\bf 15}_1^P$ & $\frac{1}{\sqrt{6}}\frac{(N_c-1)\sqrt{(-S)(N_c+2S+5)}}{(N_c+1)(N_c+5)}$ & $-\frac{1}{\sqrt{6}}\frac{(N_c-1)\sqrt{(-S)(N_c+2S+5)}}{(N_c+1)(N_c+5)}$ \\ \hline
  ${\bf \overline{6}}_1^P-{\bf 15}_1^P$ & $-\frac{1}{\sqrt{2}}\frac{1}{N_c+1}\sqrt{\frac{(-S)(2-S)(N_c-1)}{N_c+5}}$ &  $\frac{1}{\sqrt{2}}\frac{1}{N_c+1}\sqrt{\frac{(-S)(2-S)(N_c-1)}{N_c+5}}$\\ \hline
${\bf 15}^{\prime P}_1-{\bf 15}_1^P$ & $\frac{\sqrt{5}}{4}\sqrt{\frac{(-S)(S+4)(N_c+5)}{(N_c^2-1)(N_c+7)}}$ & $-\frac{\sqrt{5}}{4}\sqrt{\frac{(-S)(S+4)(N_c+5)}{(N_c^2-1)(N_c+7)}}$ \\ \hline
\end{tabular}
\caption[Matrix elements of $SU(3)$ breaking operators $B_1$, $B_2$]{Matrix elements of two of the $SU(3)$ breaking operators at order $\epsilon$, $B_1$ and $B_2$.  The results for the ${\bf 15}_1$ have been expanded to order $1/N_c$.  $I$ is the isospin and $S$ is the strangeness of a particular state.  Note that all matrix elements not diagonal in $I$ and $S$ vanish.  For example, the ${\bf 3}_1$ and ${\bf \overline{6}}_1$ mix only for $S=-1, I=1/2$.

The factor $A$ depends on both strangeness and isospin: 
$A=4I^2+4I-S^2$.}

\end{table}
%\end{landscape}

\newpage
\section{Matrix Elements of $SU(3)$ Breaking Operator $B_3$}

\begin{table}[h]
\renewcommand{\arraystretch}{2.2}
\begin{tabular}{|| c | c | }
  \hline
  &   $B_3$  
  \\ 
 &   $\frac{1}{N_c}d^{8ab}g^{ia}G_c^{ib}$  
 \\ \hline 
 ${\bf 3}_0^P$   &  $\frac{(2+3S)N_c^2+(24+18S)N_c+(70+39S)}{8\sqrt{3}N_c(N_c+1)(N_c+5)}$ \\ \hline
  ${\bf 3}_1^P$ & $-\frac{(2+3S)N_c^2+(24+18S)N_c+(70+39S)}{32\sqrt{3}N_c(N_c+1)(N_c+5)}$
     \\ \hline
 ${\bf \bar 6}^P_1$   & $-\frac{(3N_c+7)(N_c+3S+1)}{48\sqrt{3}N_c(N_c+1)}$ \\ \hline
${\bf 15}_0^P$ &  $-\frac{4N_c^2+N_c(14S+16+A)+46S+12+A}{64\sqrt{3}N_c(N_c+5)}$  \\ \hline
${\bf 15}_1^P$ & $-\frac{24N_c^4+(198S+3A+16)N_c^3}{96\sqrt{3}N_c^4}$ \\ \hline
${\bf 15}_2^P$ &  $-\frac{8N_c^2+N_c(22S+24-A)-26S-32-7A}{128\sqrt{3}N_c(N_c-1)}$  \\ \hline
${\bf 15}^{\prime P} _1$ & $-\frac{6N_c^2+N_c(44+27S)+(14+69S)}{96\sqrt{3}N_c(N_c+7)}$  \\ \hline
 ${\bf 3}_0^P-{\bf 15}_0^P$ &  $\frac{3N_c^2+4N_c-7}{32\sqrt{3}N_c(N_c+5)}\sqrt{\frac{(-3S)(N_c+2S+5)}{N_c^2-1}}$   \\ \hline
 ${\bf 3}_1^P-{\bf \overline{6}}_1^P$ &  $-\frac{3N_c^3+29N_c^2+93N_c+99}{128\sqrt{3}N_c(N_c+1)(N_c+3)}\sqrt{\frac{N_c-1}{(N_c+5)(N_c+3)}}$ \\ \hline
 ${\bf 3}_1^P-{\bf 15}_1^P$ &  $-\frac{(3N_c^2+20N_c+9)\sqrt{(-2S)(N_c+2S+5)}}{32\sqrt{3}N_c(N_c+1)(N_c+5)}$ \\ \hline
  ${\bf \overline{6}}_1^P-{\bf 15}_1^P$ & $\frac{N_c+3}{48\sqrt{3}N_c(N_c+1)}\sqrt{\frac{(N_c-1)(-6S)(2-S)}{N_c+5}}$\\ \hline
${\bf 15}^{\prime P}_1-{\bf 15}_1^P$ &  $-\frac{\sqrt{5}}{96}\frac{N_c+3}{N_c}\sqrt{\frac{(N_c+5)(2-S)}{(N_c+7)(N_c^2-1)}}$ \\ \hline
\end{tabular}
\caption[Matrix elements of $SU(3)$ breaking operator $B_3$]{Matrix elements of  the third $SU(3)$ breaking operator at order $\epsilon$, $B_3$.  

As before, $A=4I^2+4I-S^2$.  The entry for ${\bf 15}_1$ has been expanded to $\mathcal{O}(\epsilon/N_c$.}
\end{table}

\chapter{Flavor $SU(3)$ Multiplets}

\section{Negative-Parity Heavy Pentaquark Multiplets for $N_c=3$}

\begin{figure}[h]
\PSbox{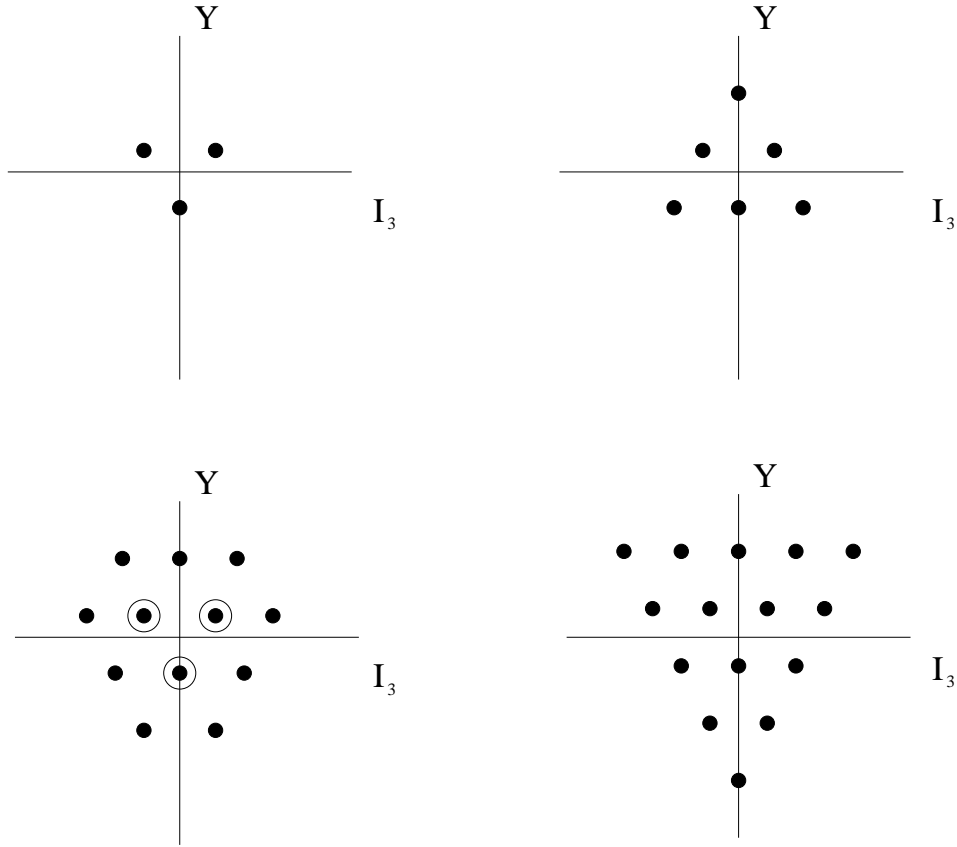 hoffset=10 voffset=30 hscale=60
vscale=60}{5.5in}{5.0in}
\caption[The four possible $SU(3)$ flavor multiplets for the negative-parity heavy pentaquarks]{The four possible $SU(3)$ flavor multiplets for the negative-parity heavy pentaquarks at $N_c=3$.  The top row shows the ${\bf 3}^P$ and the ${\bf \bar 6}^P$; the bottom row shows the ${\bf 15}^P$ and the ${\bf 15}^{\prime P}$.}
\label{fig:su3}
\end{figure}

\section{Extension to $N_c>3$}

\begin{figure}[h]
\PSbox{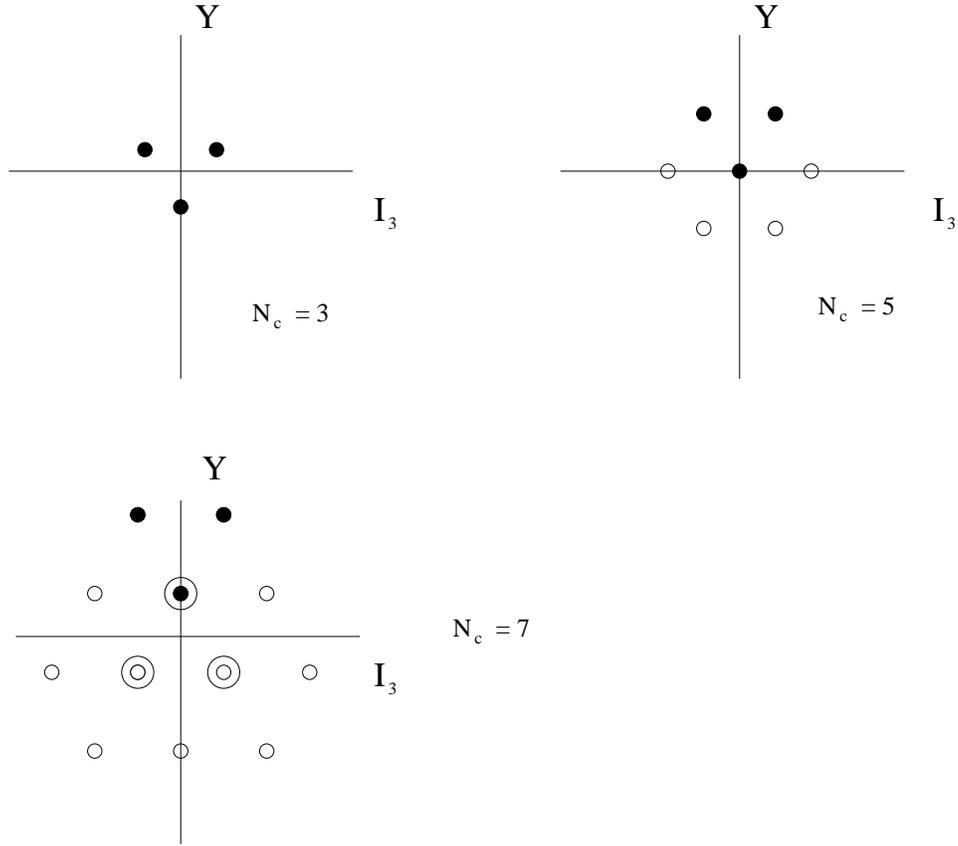 hoffset=10 voffset=30 hscale=60
vscale=60}{5.5in}{5.0in}
\caption[Extension of the ${\bf 3}$ to $N_c>3$]{Extension of the ${\bf 3}$ to $N_c>3$.  The ${\bf3}$, shown at top left for $N_c=3$ and written in Dynkin index notation as $(1,0)$, becomes $(1, \frac{N_c-1}{2})$ for arbitrary odd $N_c>3$. The first two cases are shown here: $N_c=5$ gives the ${\bf 8}$ representation, $(1,1)$, and $N_c=7$ gives the ${\bf \overline{15}}$, $(1,2)$.  In each case, the states of interest are shown in black.  For all $N_c$, they have the same isospin and strangeness values, $(I,S)=(1/2, -1)$ and $(0, -2)$.  Their hypercharges are $\frac{N_c-2}{3}$ and $\frac{N_c-5}{3}$, respectively.}
\label{fig:extension}
\end{figure}

\chapter[Gravitation Between Strings]{The Long Range Gravitational Potential Energy Between Strings}

In classical 2+1 dimensional general relativity, a point mass at rest does not result in a curved space-time away from the location of the particle. Instead, the space-time remains flat, but with a deficit angle cut out; the size of that angle is proportional to the mass of the particle \cite{djt}. This corresponds to a curvature singularity at the location of the particle. Hence in $2+1$ dimensional space-time there is no classical force between two point masses. Similarly, in $3+1$ dimensional general relativity, an infinitely long straight string, characterized only by its tension, leaves the exterior space-time flat, and the classical force between two
parallel infinitely long straight strings vanishes \cite{v}. The main purpose of this paper is to calculate the leading quantum mechanical long range force, or, equivalently potential energy, between such strings. Towards the end of this paper, we will also consider contributions to the long range force that would arise if, in addition to the massless graviton, there were
a massless scalar in the bulk. We then briefly discuss the generalization of this to other co-dimension two objects ({\it i.e.}, $p$-branes in $p+2+1$ dimensional space-time). In models with two large extra dimensions, this potential between three-branes may be relevant for cosmological quintessence \cite{cds}.

The action for the two-string system is taken to be
\begin{equation}
\label{eq1} 
S=S_b+S_1+S_2.
\end{equation}
The bulk action, $S_b$, is the usual Einstein-Hilbert action
\begin{equation}
\label{bulk}
S_b=-2M^{(n-2)}\int d^nx~\sqrt g~ R,
\end{equation}
where $R$ is the curvature scalar, and Newton's constant $G_N$ is related to the
mass $M$ by $G_N=1/(32 \pi M^2)$. Even though the long range potential is finite, it is convenient to regulate the theory using dimensional regularization, and in Eq. (\ref{bulk}) $n=4-\epsilon$. For the two string actions, $S_i$, we take
\begin{equation}
\label{brane}
S_i=-\tau_i \int d^nx~ \sqrt{ g^{(i)}}~ \delta^{(2)} (\vec x-\vec x_i),
\end{equation}
where $g^{(i)}$ is the induced metric on the world-sheet of string $i$. Note that in $n$ dimensions the string world-sheets have dimension $n-2$, so they are still co-dimension two objects. 
We have chosen to align the strings along the $1$ axis; the separation between the two strings is $\vec a=\vec x_1-\vec x_2$. Indices that go over the 4 space-time coordinates $0,1,2,3$ ($n$ space-time coordinates in $n$ dimensions) are denoted by capital Roman letters; those that just go over the 2 space-time coordinates of the string world-sheet $0,1$ are denoted by Greek letters. Finally, indices that take on values in the two spatial directions perpendicular to the strings are denoted by lower case Roman letters, and vectors in the $2,3$ plane are denoted with arrows. We align the local space-time coordinates on the string world surfaces with those of the bulk space-time, so the components of the induced metric tensor are the same as those of the bulk metric but restricted to the $0,1$ values of the indices, {\it i.e.},  $g^{(i)}_{\alpha \beta}=g_{\alpha \beta}$.

Expanding the gravitational field as\footnote{Here $\eta=diag[-1,1,1,1]$ is the usual flat space-time metric tensor.} 
\begin{equation}
g_{MN}=\eta_{MN} + h_{MN}/M^{(n /2 -1)},
\end{equation}
we determine the leading quantum contribution to the potential between the two strings by computing one-loop Feynman diagrams with vertices that follow from the action in Eq. (\ref{eq1}). This is similar to the computation of the quantum correction to the Newtonian potential between point masses\footnote{ There is some ambiguity in precisely how the potential is defined. This issue is less severe for strings since the classical potential vanishes.} in four space-time dimensions \cite{dhl}. The main difference between the string
and point mass cases is that for strings the classical force vanishes; hence our computation gives the leading contribution to the force instead of a small correction.  

't Hooft and Veltman calculated the infinite part of the one-loop gravitational effective action \cite{hv}. We adopt the same background field gauge fixing, so the momentum space propagator for the canonically normalized graviton field is
\begin{equation}
D_{AB,CD}(x-y)=P_{AB,CD}D(x-y), 
\end{equation}
where $D(x-y)$ is the usual scalar propagator with Fourier transform $D(q)=-i/(q^2-i \epsilon)$ and 
\begin{equation}
P_{AB,MN}={1 \over 2}[\eta_{AM} \eta_{BN} +\eta_{AN} \eta_{BM}-{2 \over n-2}\eta_{AB} \eta_{MN}].
\end{equation}
It is convenient to use their gauge fixing convention, because then the contribution of some of the Feynman diagrams to
the quantum force between strings can be deduced from their work. Unless explicitly stated otherwise, indices on $h$ and $P$ are raised and lowered with the flat space metric tensor $\eta$. 

\begin{figure} [h] 
\PSbox{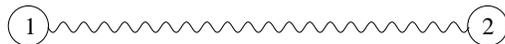 hoffset=110 voffset=10 hscale=40 vscale=40}{6.8 in}{1.0 in}
\caption[Classical contribution to the potential]{Classical contribution to the potential.  The numbers 1 and 2 represent the two string world-sheets.}
\label{fig:fig1}
\end{figure}

In this paper, we treat the tensions as small compared with $M^2$, and only keep the terms in the potential proportional to the product of the two tensions $\tau_1 \tau_2$, neglecting terms suppressed by additional powers of $G_N\tau_i$.  Using perturbation theory, it is easy to understand why the part of the classical potential proportional to $\tau_1 \tau_2$ vanishes. It comes from the Feynman diagram in Figure \ref{fig:fig1}. Using 
\begin{equation}
\sqrt {g^{(i)}}=1+h_\alpha^\alpha/2M^{(n/2-1)}+ h_{\alpha}^{\alpha}~h_{\beta}^{\beta}/8M^{(n-2)}-h_{\beta}^{\alpha}~h_{\alpha}^{\beta}/4M^{(n-2)}+.... .  , 
\end{equation}
it follows that this diagram is proportional to $P_{\alpha}^ {\alpha},^{\mu}_{ \mu}=\eta^{\alpha \beta} \eta^{\mu \nu}P_{\alpha \beta,\mu \nu}=0$.

In background field gauge, one decomposes the graviton field into quantum and classical pieces: $h= \bar h + \tilde h$, where the bar denotes the classical part and the tilde the quantum part. The leading quantum correction occurs at one-loop. The quantum fields are contracted to make the propagators that occur in the loop; the classical fields are contracted for the other propagators.  (In the figures, classical gravitons are drawn as wavy lines, quantum gravitons as curly lines.) 

\begin{figure}
\PSbox{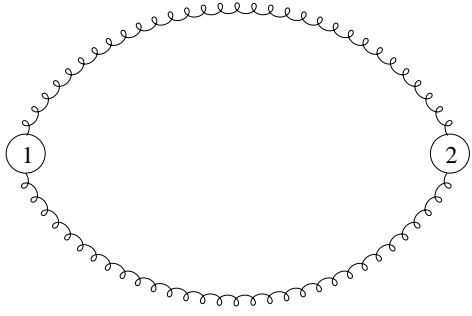 hoffset=130 voffset=10 hscale=40
vscale=40}{6.8in}{2.0in}
\caption[Feynman diagram determining the one-loop contribution from terms localized on the string
 world-sheets]{Feynman diagram that determines the one-loop contribution to the potential from terms localized on the string
 world-sheets that are quadratic in the graviton field.}
\label{fig:fig2}
\end{figure}

First we consider Figure \ref{fig:fig2}. The coupling of the gravitons to the string world-sheet comes from the quadratic terms in the expansion of the square root of the induced metric in $h$. The contribution to the effective action that results from this Feynman diagram is

\begin{equation}
\label{act1}
i\Delta S_{eff}=-{\tau_1 \tau_2 \over M^4}\left[{P^{\alpha}_{\alpha},^{\beta}_{\beta}P^{\lambda}_{\lambda},^{\delta}_{\delta} \over 32}-{P^{\alpha}_{\alpha},^{\lambda}_{\beta}P^{\delta}_{\delta},^{\beta}_{\lambda}\over 8}+{P^{\alpha}_{\beta},^{\lambda}_{\delta}P^{\beta}_{\alpha},^{\delta}_{\lambda}\over 8}\right]\int d^2x_1d^2x_2D(x_1-x_2)^2.
\end{equation}
Eq. (\ref{act1}) is evaluated using 
\begin{equation}
\int d^2x_1d^2x_2D(x_1-x_2)^2=-i{LT \over a^2}\int{d^2k \over (2\pi)^2}{K_0(k)^2 \over (2\pi)^2}=-i{LT \over a^216 \pi^3},
\end{equation}
where $K_0(k)$ is a Bessel function of imaginary argument, the integrals go over the world-sheets of the two strings, and
\begin{equation}
P^{\alpha}_{\alpha},^{\beta}_{\lambda}=0, ~~~~~~~~~~~~~P^{\alpha}_{\beta},^{\lambda}_{\delta}P^{\beta}_{\alpha},^{\delta}_{\lambda}=2.
\end{equation}
The effective action can be interpreted as minus the potential energy times the time, $\Delta S_{eff}=-\Delta U~T$.
Putting these results together, we find that the contribution to the potential energy per unit string length from this diagram is
\begin{equation}
\Delta U/L={\tau_ 1\tau_2 \over 16 \pi^3 a^2M^4}\left(-{1 \over 4}\right).
\end{equation}

\begin{figure} [h]
\PSbox{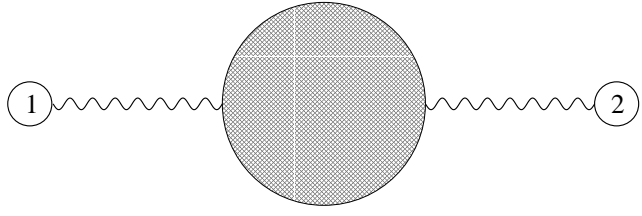 hoffset=100 voffset=10 hscale=45
vscale=45}{6.8in}{1.5in}
\caption[Feynman diagrams that give the contribution to the potential from gravitational self interactions]{Feynman diagrams that give the contribution to the potential from gravitational self interactions. The shaded circle includes gravitons and ghosts in the loop.}
\label{fig:fig3}
\end{figure}

Next consider the diagrams in Figure \ref{fig:fig3}. 't Hooft and Veltman  \cite{hv} found that the divergent part of the one-loop gravitational effective action for pure Einstein gravity is
\begin{equation}
\label{div}
S_{1loop}^{eff}=-{M^{n-4} \over 8 \pi^2 (n-4)}\int d^n x \left({1 \over 120} R^2 +{7 \over 20} R_{AB}R^{AB}\right).
\end{equation}
 From this we can deduce the insertion appropriate for the shaded circle in Figure \ref{fig:fig3} by expanding out the curvature tensor to linear order in the gravitational field.  We get
\begin{equation}
\label{r2}
R^2=\left[(\partial ^2 h^L_L)(\partial^2h^M_M)-2 (\partial^2 h^L_L)( \partial_G \partial_E h^{GE})+(\partial_K \partial_N h^{KN})( \partial_G \partial_E h^{GE})\right]/M^{(n-2)},
\end{equation}
and
\begin{align}
\label{ri2}
&R_{MK}R^{MK}=\left[{1 \over 4}(\partial_K \partial_M h^L_L)( \partial^K \partial ^M h^N_N)-(\partial_K \partial_M h^L_L)(\partial ^K \partial^N h_N^M) +{1 \over 2} (\partial _K \partial_M h^L_L)( \partial^2 h^{KM} )\right.
\nonumber \\
& +{1 \over 2}(\partial_K \partial_L h^{LM})( \partial^K \partial^N h_{MN}) +{1 \over 2} (\partial_K \partial_L h^{LM})(\partial_N \partial _M h^{KN})- (\partial_K \partial_L h^L_M)( \partial^2 h^{KM} )\nonumber \\
&\left. +{1 \over 4 }(\partial^2 h_{KM})( \partial^2h^{KM} )\right]/M^{(n-2)}.
\end{align}
The contribution from the one loop diagram in Figure \ref{fig:fig3} is deduced by inserting in the momentum space vertex associated with the the action in Eq. (\ref{div}) an additional factor of $q^{2(n/2-2)}=1-\epsilon \ln(q^2)/2+...$. Without this factor, the contribution to the long range force between strings would vanish. It is the finite nonanalytic part of the effective action, not the divergent part, that is actually responsible for the long range force. The momentum space integral that must be done is then
\begin{equation}
-{1 \over n-4}\int {d^2 q \over (2 \pi)^2} \exp(i\vec q \cdot \vec a)({\vec q}^2)^{(n/2-2)}= {1 \over 2 \pi a^2}.
\end{equation}
Putting these results together, the contribution from the diagrams in Figure \ref{fig:fig3} to the long range potential energy per unit length between strings is
\begin{align}
\Delta U/L&={\tau_ 1\tau_2 \over 16 \pi^3 a^2M^4}\left({1 \over 120}\left[-2+2-{1 \over 2}\right]+{7 \over 20} \left[-{1 \over 2}+1-{1 \over 2}-{1 \over 4}-{1 \over 4}+{1 \over 2}-{1 \over 4}\right]\right) \nonumber  \\
&={\tau_ 1\tau_2 \over 16 \pi^3 a^2M^4}\left(-{11 \over 120}\right).
\end{align}
The successive terms in the square brackets are the contributions of the corresponding terms in the square brackets of Eqs. (\ref{r2}) and (\ref{ri2}).

\begin{figure}
\PSbox{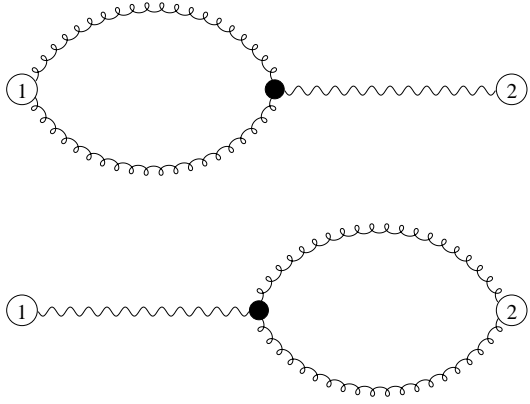 hoffset=120 voffset=10 hscale=35
vscale=35}{6.8in}{2.5in}
\caption[Contribution to the one loop potential from the three-graviton vertex from the Einstein-Hilbert action]{Contribution to the one loop potential that arises from the three-graviton vertex from the Einstein-Hilbert action.}
\label{fig:fig4}
\end{figure}

Next we consider the contribution to the long range force from the Feynman diagrams in Figure \ref{fig:fig4}. For this we need the bulk three-graviton vertex from the Einstein-Hilbert term. It comes from expanding the action in Eq. (\ref{bulk}) to cubic order in $h$, yielding
\begin{equation}
S_{3h}=-{2 \over M^{(n/2-1)}}\int d^n x~ {\cal L}_1 +{ \cal L}_2 + {\cal L}_3,
\end{equation}
where ${\cal L}_i$ is the part that comes from expanding the curvature tensor to order $i$ in $h$. Explicitly,
\begin{equation}
{\cal L}_1={1 \over 8} h^A_A h^B_B \partial^2 h^L_L -{1 \over 8} h^A_A h^B_B \partial_K \partial_N h^{KN} -{1\over 4} h^B_A h^A_B \partial ^2 h^L_L +{1\over 4} h^B_A h^A_B \partial _K \partial_N h^{KN},
\end{equation}
\begin{align}
&{\cal L}_2 ={1 \over 2}h^A_A (\partial_L h^{EL})(\partial_M h^M_E)-{1 \over 2} h^A_A (\partial_L h^{EL}) (\partial_E h^M_M)  +{1 \over 8} h^A_A (\partial^E h^L_L) (\partial_E h^M_M) \nonumber \\
&-{3 \over 8} h^A_A (\partial_L h^{EM})(\partial^L h_{EM}) +{1 \over 4} h^A_A (\partial_M h^E_L)(\partial^L h^M_E) -{1 \over 2}h^{NL} h^A_A \partial^2 h_{NL}+h^{NL}h^A_A\partial_M \partial_L h^M_N \nonumber \\
&-{1 \over 2} h^{NL}h^A_A \partial_N \partial_L h^M_M ,
\end{align}
and
\begin{align}
&{\cal L}_3 =-h^G_E(\partial^L h^E_L)(\partial_M h^M_G)+h^G_E (\partial_L h^{LE})(\partial_G h^M_M)-{1 \over 4}h^G_E(\partial^E h^L_L)(\partial_G h^M_M)\nonumber \\
&-{1 \over 2}h^G_E(\partial_L h^{EM})(\partial_M h^L_G)+{3 \over 2}h^G_E(\partial_L h^{EM})(\partial^L h_{MG})-h^G_E(\partial_L h^{EM})(\partial_G h^L_M)\nonumber \\
&+{3 \over 4}h^G_E(\partial^E h_{ML})(\partial_G h^{ML})-2h^G_E(\partial_G h^{LE})(\partial_M h^M_L) 
+h^G_E(\partial_G h^{LE})(\partial_L h^M_M)\nonumber \\
&+h^G_E(\partial^L h^E_G)(\partial_M h^M_L)-{1 \over 2}h^G_E(\partial^L h^E_G)(\partial_L h^M_M) +h^N_Bh^{BL} \partial^2 h_{LN} -2 h^N_B h^{BL}\partial_M \partial_L h^M_N \nonumber \\
&+h^N_B h^{BL}\partial_N \partial_L h^M_M +h^{NL}h^{MK}\partial_K \partial_M h_{NL}-h^{NL}h^{MK}\partial_K \partial_L h_{NM}.
\end{align} 
The integral needed to compute Figure \ref{fig:fig4} is
\begin{equation}
\int d^2 x_1 d^2 x_2 d^4 x (\partial_x^2 D(x_1-x))D(x-x_2)^2={LT \over 16 \pi^3 a^2},
\end{equation}
and we find that it gives the following contribution to the gravitational potential per unit length:
\begin{align}
\label{r3} 
&\Delta U/L={\tau_ 1\tau_2 \over 16 \pi^3 a^2M^4}\left( \left[0+0+1-{1 \over 2}\right]+\left[-{1 \over 6}+0+0+{3 \over 4}-{1 \over 12}+0+{1 \over 3}+0\right] \right.\nonumber \\
&+\left. \left[0+0+0+0+0+0-{1 \over 2}+0+0+1-1+0+0+0-{1 \over 3}+0\right ]   \right),
\end{align}
In Eq. (\ref{r3}) the three square brackets contain the contributions from the three Lagrange densities ${\cal L}_1$, ${\cal L}_2$, and ${\cal L}_3$ respectively, and each successive term in these square brackets represents the contribution of the corresponding term in the Lagrange density. Summing them up, we get
\begin{equation}
\Delta U/L={\tau_ 1\tau_2 \over 16 \pi^3 a^2M^4}\left( {1 \over 2} \right),
\end{equation}
for the contribution from diagrams which contain a three graviton vertex from the Einstein-Hilbert action. 

\begin{figure}
\PSbox{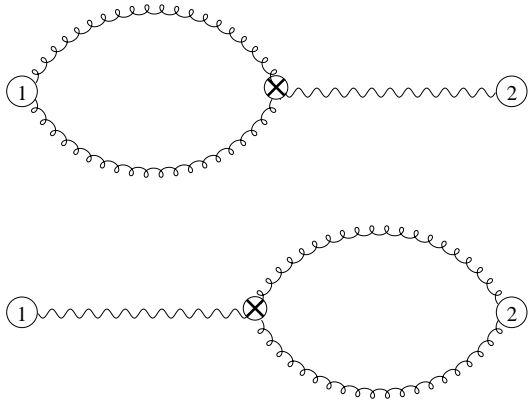 hoffset=120 voffset=10 hscale=35
vscale=35}{6.8in}{2.5in}
\caption[One loop contribution to the potential from the three-graviton vertex from gauge fixing]{One loop contribution to the potential that arises from the three-graviton vertex from gauge fixing.}
\label{fig:fig5}
\end{figure}

In the background field gauge, the gauge fixing term also contributes to the $\bar h \tilde h \tilde h$ vertex. Using the definition $\bar g_{AB}= \eta_{AB} +\bar h_{AB}/M^{(n/2-1)}$, the gauge fixing term is \cite{hv}
\begin{equation}
\label{gf}
S_{gf} =- \int d^n x \sqrt{ \bar g}\left(D_N \tilde h^N_M -{1 \over 2} D_M \tilde h^L_L\right)\left(D_S \tilde h^{MS} -{1 \over 2} D^M \tilde h^S_S \right),
\end{equation} 
where in Eq. (\ref{gf}) indices are raised and lowered with the classical metric $\bar g$ and the covariant derivative is with respect to this metric. Expanding to linear order in $\bar h$, the above becomes
\begin{equation}
\label{gf1}
S_{gf} =-\int d^n x {1 \over 2M^{(n/2-1)}} \left(\bar h^A_A (\partial_N \tilde h_M^N)(\partial_S \tilde h^{MS})+...\right)+...,
\end{equation} 
where the ellipses in the brackets denote other terms linear in $\bar h$, and the ellipses outside of the brackets denote terms higher order in $\bar h$. Only the term explicitly displayed in equation (\ref{gf1}) contributes at one loop; the other terms linear in $\bar h$ (represented by the ellipses inside the brackets) each give zero. We find that the contribution of Figure \ref{fig:fig5} to the long range potential is

\begin{equation}
\Delta U/L={\tau_ 1\tau_2 \over 16 \pi^3 a^2M^4}\left( -{1 \over 12} \right).
\end{equation}

All other possible one loop contributions vanish. For example, there is a cubic coupling of $h_{\mu \nu}$ on the brane from expanding the induced metric to that order. The one loop graph formed from this coupling  vanishes since $P^{\alpha}_{\alpha},^{\beta}_{\lambda}=0$. 

%\begin{figure} [b]
%\PSbox{fig6.eps hoffset=120 voffset=5 hscale=40
%vscale=40}{6.8in}{1.8in}
%\caption{One loop contribution from cubic graviton coupling on the string world sheet. There is %a similar diagram where the cubic coupling is on the world sheet of string 2}
%\label{fig:fig6}
%\end{figure}

So far we have not included the degrees of freedom that correspond to transverse fluctuations of the strings. However, they must exist, by reparametrization invariance and general covariance. These fluctuations are characterized by scalar fields $\phi^a_{(i)}$ which are localized on the world-sheet of string $i$. The terms in the string actions (\ref{brane}) involving the fields $\phi^a_{(i)}$ are deduced from the dependence of the induced metric\footnote{See, for example, \cite{s1}.} on them, 
\begin{equation}
g^{(i)}_{\mu \nu}= g_{\mu \nu} + g_{a b}(\partial_{\mu} \phi_{(i)}^a)(\partial_{\nu} \phi_{(i)}^b).
\end{equation}
Expanding the square root of the determinant of the above induced metric yields a coupling of $h_{ab}$ to the scalar fields. However, the graph with a $\phi$ loop vanishes in dimensional regularization since it 
is proportional to $\int d^{(n-2)}k=0$.

%\begin{figure} 
%\PSbox{fig7.eps hoffset=100 voffset=10 hscale=40
%vscale=40}{6.8in}{1.5in}
%\caption{Contribution to the long range potential from the graviton coupling to the scalar %fields that describe fluctuations of the string world sheets. Similar diagram occurs where the %scalar fields couple to the world sheet of string 2.}
%\label{fig:fig7}
%\end{figure}

Summing the various one loop contributions to gravitational potential energy between strings gives
\begin{equation}
U/L={\tau_ 1\tau_2 \over 16 \pi^3 a^2M^4}\left( {3 \over 40} \right)={24G_N^2\tau_1\tau_2 \over5 \pi a^2}.
\end{equation}
The above equation is the main result of this paper. It gives a repulsive gravitational force between the strings at large distances. 

From the effective field theory point of view, it is possible that the tree level effects of higher dimension operators are of the same size as the one loop pieces we have calculated, but this turns out not to be the case.  Nontrivial operators localized on the string world sheet with less than two derivatives are forbidden by general covariance. Furthermore, we know that many operators do not contribute to the long range force. Consider, for example, adding to the string world-sheet actions the following two-derivative term:
\begin{equation}
\delta S_i = \lambda_i \int d^4 x \sqrt{ g^{(i)}}~ R~ \delta^{(2)} (\vec x-\vec x_i),
\end{equation}
where the $\lambda_i$ are dimensionless couplings. Classically there is no  contribution to the long range force between the branes linear in these couplings. At this order it only gives local effects proportional to $\delta^{(2)}(\vec a)$ or derivatives of this delta function. Similar remarks hold for operators in the bulk that are quadratic in the curvature tensor. We will not attempt a complete analysis of the tree level effects from higher dimension operators; however, there is no tree level contribution to the potential that is as important at large $a$ as the one loop piece we have calculated.

Effects similar to what we have computed occur for $p$-branes in a space-times of dimension $p+3$. Assuming that gravity is the only massless degree of freedom in the bulk, there will be a long range contribution to the potential per unit $p$-brane volume proportional to $G_N^2 \tau_1 \tau_2/a^{p+1}$ from one loop quantum effects. If there are other massless degrees of freedom in the bulk, these will also contribute to the long range force. Consider, for example, a scalar field theory with space-time dimension $p+3$ and two parallel $p$-branes. Neglecting gravity, the action for this system is taken to be
\begin{equation}
\label{act} 
S=S_b+S_1+S_2,
\end{equation}
where the bulk action, $S_b$, comes from the massless Klein Gordon theory:
\begin{equation}
S_b=-{1 \over 2}\int d^{p+3}x~\partial_M \chi \partial^M \chi,
\end{equation}
and the brane actions are
\begin{equation}
S_i=-{\lambda_i \over 2}\int d^{p+3}x~ \chi^2 \delta^{(2)} (\vec x-\vec x_i).
\end{equation}

\begin{figure} 
\PSbox{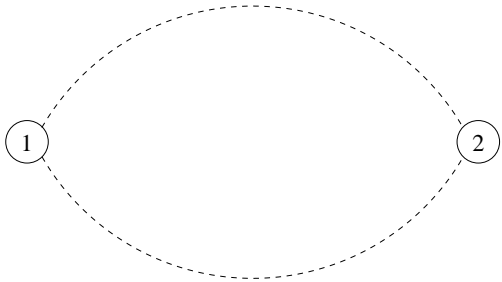 hoffset=100 voffset=10 hscale=40
vscale=40}{6.8in}{2.0in}
\caption[One loop contribution to the potential from massless bulk scalar with brane mass terms]{One loop contribution to the potential from massless bulk scalar with brane mass terms.}
\label{fig:fig8}
\end{figure}

Because of the $\chi \rightarrow - \chi$ symmetry there is no tree level force between the branes from $\chi$ exchange. Assuming that the couplings $\lambda_i$ are small and neglecting effects higher order in these coupling constants, the one loop diagram in Figure \ref{fig:fig8} gives the long range potential\footnote{For work in string theory 
on the force between branes see \cite{bs}.}
\begin{equation}
\label{spot}
U/V=-{\lambda_1 \lambda_2  \Gamma({p \over 2}+{1 \over 2})^2 \over a^{p+1}p 2^{p+4} \pi^{({p \over 2}+2)}\Gamma({p\over 2})}.
\end{equation}
If the couplings $\lambda_1$ and $\lambda_2$ have opposite signs, this potential is repulsive. It can be natural for the scalar to have brane mass terms but no bulk mass. For example, $\chi$ could be the Goldstone boson associated with a global symmetry that is spontaneously broken in the bulk but explicitly broken on the branes. At higher order the couplings $\lambda_i$ become subtraction point dependent \cite{gw}.

Let's focus on the case of three-branes in six dimensions. If the two dimensions perpendicular to the branes are compact but large extra dimensions of the type that has been suggested to be related to the hierarchy puzzle \cite{add}, then the potential in Eq. (\ref{spot}) has the right form to be suitable for cosmological quintessence\footnote{ This is similar to the proposal in \cite{abrs}.}. The scalar field has mass dimension two, so the parameters $\lambda_i$ are dimensionless. The separation between the branes is related to the scalar fields that characterize the brane fluctuations. Assuming the two compact extra dimensions are flat,\footnote{The physics that determines the size of the compact two extra dimensions is assumed to be unrelated to the potential generated by $\chi$ loops.} the action for the scalar fields that characterize the fluctuations of the 3-brane world-sheets is
\begin{equation}
S_{fluct}=-\tau_1 \int d^4 x_1 {1 \over 2}\partial_{\mu}\phi^a_{(1)} \partial^{\mu} \phi^a_{(1)}-\tau_2 \int d^4 x_2 {1 \over 2}\partial_{\mu}\phi^a_{(2)} \partial^{\mu} \phi^a_{(2)} +....~.
\end{equation}
The repeated index $a$, which takes on the values $1,2$, is summed over. From the four dimensional effective field theory point of view, this action becomes
\begin{equation}
S^{eff}_{4dim}=-\int d^4x~ { \tau_1+\tau_2 \over 2}\partial_{\mu} \phi^a_{cm} \partial^{\mu} \phi^a_{cm}+{\tau_r \over 2} \partial_{\mu} \phi^a_{rel}\partial^{\mu} \phi^a_{rel}+...,
\end{equation}
where $\tau_r=\tau_1 \tau_2/(\tau_1+\tau_2)$ is the reduced tension, $\phi^a_{rel}= \phi^a_{(1)}- \phi^a_{(2)}$ and $\phi^a_{cm}=(\tau_1 \phi^a_{(1)}+\tau_2 \phi^a_{(2)})/(\tau_1+\tau_2)$. The separation between the branes $\vec a $ is the vacuum expectation value of $\vec \phi_{rel}$, and the canonically normalized four dimensional field associated with the separation between the branes is $\vec \phi=\sqrt \tau_r \vec \phi_{rel}$. The potential for this scalar field is of the form $ U/V \sim -\lambda_1 \lambda_2 \tau_r^2/  (\vec \phi ^2)^2 \sim M_W^8/(\vec \phi^2)^2$, when the brane tensions are of order the weak scale\footnote{With tensions this large it is no longer a good approximation to neglect the deficit angles associated with these three-branes.}. In cosmological quintessence the scalar field today is of order the Planck mass; this corresponds to a separation between branes of order the size of the compact space ({\it i.e.}, of order a millimeter). Clearly, the impact of the physics that stabilizes the compact dimensions  \cite{stab} has to be taken into account before the true physical significance of this potential can be ascertained.

\end{document}